\documentclass[11pt]{article}
\usepackage{amsmath,amsfonts}
\usepackage{epsfig}

\begin{document}

\title{\textbf{The influence of the Gribov copies on the gluon and ghost
propagators in Euclidean Yang-Mills theory in the maximal Abelian
gauge}}
\author{\textbf{M.A.L. Capri}\thanks{
marcio@dft.if.uerj.br} \ , \textbf{V.E.R. Lemes}\thanks{
vitor@dft.if.uerj.br} \ , \textbf{R.F. Sobreiro}\thanks{
sobreiro@uerj.br} \and \textbf{S.P. Sorella}\thanks{
sorella@uerj.br}{\ }\footnote{Work supported by FAPERJ, Funda{\c
c}{\~a}o de Amparo {\`a} Pesquisa do Estado do Rio de Janeiro,
under the program {\it Cientista do Nosso Estado},
E-26/151.947/2004.} \ , \textbf{R. Thibes}\thanks{
thibes@dft.if.uerj.br} \\
\\
\textit{UERJ, Universidade do Estado do Rio de Janeiro} \\
\textit{Rua S{\~a}o Francisco Xavier 524, 20550-013 Maracan{\~a}} \\
\textit{Rio de Janeiro, Brasil}} \maketitle

\begin{abstract}
The effects of the Gribov copies on the gluon and ghost propagators are
investigated in $SU(2)$ Euclidean Yang-Mills theory quantized in the maximal
Abelian gauge. The diagonal component of the gluon propagator displays the
characteristic Gribov type behavior. The off-diagonal component of the gluon
propagator is found to be of the Yukawa type, with a dynamical mass
originating from the dimension two condensate $\left\langle A_{\mu
}^{a}A_{\mu }^{a}\right\rangle $, which is also taken into account. Finally,
the off-diagonal ghost propagator exhibits infrared enhancement.
\end{abstract}

\newpage

\section{Introduction}

Among the class of covariant gauges, the maximal Abelian gauge \cite
{'tHooft:1981ht,Kronfeld:1987vd,Kronfeld:1987ri} displays several
interesting features. This gauge is suitable for the study of the dual
superconductivity mechanism for color confinement \cite{scon}, according to
which Yang-Mills theories in the low energy region should be described by an
effective Abelian theory \cite
{Ezawa:bf,Suzuki:1989gp,Suzuki:1992gz,Hioki:1991ai} in the presence of
monopoles. A dual Meissner effect arising as a consequence of the
condensation of these magnetic charges might give rise to quark confinement.
Here, the Abelian configurations are identified with the diagonal components
$A_{\mu }^{i}$, $i=1,...,N-1$, of the gauge field corresponding to the $%
\left( N-1\right) $ generators of the Cartan subgroup of $SU(N)$. Moreover,
the remaining off-diagonal components $A_{\mu }^{a}$, $a=1,...,N^{2}-N$,
corresponding to the $\left( N^{2}-N\right) $ off-diagonal generators of $%
SU(N)$, are expected to acquire a mass through a dynamical mechanism, thus
decoupling at low energies. \newline
\newline
The maximal Abelian gauge can be formulated on the lattice \cite
{Kronfeld:1987vd,Kronfeld:1987ri}, a feature which has made possible to
investigate the gluon propagator by numerical simulations which, in the case
of $SU(2)$, have reported an effective off-diagonal gluon mass of
approximately $1.2GeV$ \cite{Amemiya:1998jz,Bornyakov:2003ee}. Another
relevant feature of the maximal Abelian gauge is its multiplicative
renormalizability to all orders of perturbation theory \cite
{Min:1985bx,Fazio:2001rm,Dudal:2004rx,Gracey:2005vu}. This property has
allowed for a study of the dynamical mass generation for off-diagonal
gluons, through the condensation of the operator\footnote{%
We remind here that, due to the nonlinearity character of the maximal
Abelian gauge, a slightly more general operator, $\left( \frac{1}{2}A_{\mu
}^{a}A_{\mu }^{a}+\alpha \overline{c}^{a}c^{a}\right) $, has to be
considered for renormalization purposes. The fields $\overline{c}^{a}$, $%
c^{a}$ denote the off-diagonal Faddeev-Popov ghosts, while $\alpha $ stands
for a gauge parameter. The operator\thinspace $\left( \frac{1}{2}A_{\mu
}^{a}A_{\mu }^{a}+\alpha \overline{c}^{a}c^{a}\right) $, introduced in \cite
{Kondo:2001nq}, is multiplicatively renormalizable to all orders \cite
{Dudal:2004rx,Gracey:2005vu,Kondo:2001tm}. The maximal Abelian gauge is
recovered in the limit $\alpha \rightarrow 0$, which has to be taken after
the removal of the ultraviolet divergences. Whenever necessary, we shall
refer to \cite{Dudal:2004rx} for the details of the renormalization aspects
of the maximal Abelian gauge as well as of the operator $\left( \frac{1}{2}%
A_{\mu }^{a}A_{\mu }^{a}+\alpha \overline{c}^{a}c^{a}\right) $.} $A_{\mu
}^{a}A_{\mu }^{a}$ \cite{Kondo:2001nq}. An effective potential for this
operator has been evaluated in analytic form in \cite{Dudal:2004rx},
providing evidence for a nonvanishing dimension two condensate $\left\langle
A_{\mu }^{a}A_{\mu }^{a}\right\rangle $. \newline
\newline
It is worth mentioning that, although the operator $A^2$ has been
proven to be multiplicatively renormalizable to all orders in the
Landau, linear covariant, Curci-Ferrari and maximal Abelian gauges
\cite {Dudal:2002pq,Dudal:2003np,Dudal:2003pe}, a satisfactory
understanding of
the aspects related to the gauge invariance of the dimension two condensate $%
\left\langle A^2 \right\rangle$ is still lacking. We refer to
\cite
{Esole:2004jd,Esole:2003ke,Slavnov:2004rz,Slavnov:2005av,Bykov:2005tx,Kondo:2005zs}
for an updated analysis of this important issue.\\\\ As other
gauges, the maximal Abelian gauge is affected by the Gribov copies
\cite{Gribov:1977wm}, whose existence stems from a general result
\cite {Singer:1978dk} on the lack of a globally well defined gauge
fixing procedure. A detailed construction of an explicit example
of a zero mode of the Faddeev-Popov operator in the maximal
Abelian gauge can be found in \cite {Bruckmann:2000xd}.
Nevertheless, a study of the influence of the Gribov copies on the
Green's functions of the theory in this gauge is still lacking.
The aim of the present paper is that of providing a first analysis
of the influence of the Gribov copies in the maximal Abelian
gauge. The need for such an investigation is motivated by the
great relevance that the Gribov copies have on the infrared
behavior of Yang-Mills theories, as one learns from the large
amount of results obtained in the Landau and Coulomb gauges \cite
{Zwanziger:1982na,Zwanziger:1988jt,Zwanziger:1989mf,Dell'Antonio:1989jn,Semenov,Dell'Antonio:1991xt,Zwanziger:1992qr, vanBaal:1991zw, Cucchieri:1996ja,Zwanziger:2002sh,Greensite:2004ke,Feuchter:2004mk,Reinhardt:2004mm}%
. Therefore, it might be useful to improve as much as possible our
understanding on the role of the Gribov copies in different gauges, as
recently discussed in the case of the linear covariant gauges \cite
{Sobreiro:2005vn}. \newline
\newline
In the following, we shall focus on the study of the gluon and ghost
propagators in the maximal Abelian gauge, with $SU(2)$ as gauge group. This
allows us to make a comparison with the results available from lattice
numerical simulations. The analysis of the Gribov copies will be done by
following Gribov's original work \cite{Gribov:1977wm}. It turns out in fact
that the construction outlined by Gribov in the case of the Landau and
Coulomb gauges can be essentially repeated and adapted to the case of the
maximal Abelian gauge. We shall begin with a discussion of the gauge fixing
condition and of the related Faddeev-Popov operator. Further, we shall
generalize to the maximal Abelian gauge Gribov's result stating that for any
field close to a horizon there is a gauge copy, close to the same horizon,
located on the other side of the horizon\footnote{%
We have found useful to collect the detailed proof of this statement in
Appendix A.} \cite{Gribov:1977wm}. We shall proceed thus by restricting the
domain of integration in the Feynman path integral to the so-called Gribov
region, \textit{i.e.} to the region in field space whose boundary is the
first Gribov horizon, where the first vanishing eigenvalue of the
Faddeev-Popov operator appears. The restriction to the Gribov region will be
implemented by means of a no-pole condition on the ghost two-point function,
as done in \cite{Gribov:1977wm}. This will lead to the introduction of the
Gribov parameter $\gamma $ and of the related gap equation, enabling us to
work out the infrared behavior of the gluon and ghost propagators. \newline
\newline
A few remarks are now in order. Considering the case of the Landau gauge, it
turns out that the restriction to the Gribov region does not eliminate all
possible copies. It has been proven in fact that Gribov copies still exist
inside the Gribov region \cite{Semenov,Dell'Antonio:1991xt,vanBaal:1991zw}.
To avoid the presence of these additional copies, a further restriction to a
smaller region, known as the fundamental modular region, should be
implemented$\footnote{%
The same conclusion holds for the Coulomb gauge.}$. Several properties of
the Gribov region as well as of the fundamental modular region have been
established in recent years \cite{Semenov,Dell'Antonio:1991xt,vanBaal:1991zw}%
. This has been possible due to the availability of an auxiliary functional%
\footnote{%
The color index $A$ runs now over all the generators of $SU(N)$, $%
A=1,...,N^2-1$.}, $\mathcal{F}[A]=\int d^{4}x$ $A_{\mu }^{A}A_{\mu }^{A}$, $%
A=1,...,N^{2}-1$, whose minimization along the gauge orbit of $A_{\mu }^{A}$
provides a characterization of both Gribov and fundamental modular region.
It turns out that the Gribov region can be defined as the set of all
relative minima in field space of this auxiliary functional, while the
fundamental modular region is identified with the set of all absolute minima
of $\mathcal{F}[A]$. Although the restriction to the Gribov region does not
eliminate all possible copies, its implementation in the Feynman path
integral can be effectively worked out \cite
{Zwanziger:1989mf,Zwanziger:1992qr}, allowing one to obtain a certain amount
of information on the infrared behavior of the gluon and ghost propagators.
Such a task appears to be considerably difficult in the case of the modular
region and, to our knowledge, it has not yet been accomplished. Here, a
finite volume Hamiltonian approach proves to be more adequate \cite
{vanBaal:1982ag,Koller:1987fq,vanBaal:1990ji}, see \cite{vanBaal:2000zc} for
a review. \newline
\newline
Concerning now the maximal Abelian gauge, it is worth noting that a suitable
auxiliary functional can be introduced also here, namely $\mathcal{R}%
[A]=\int d^{4}x$ $A_{\mu }^{a}A_{\mu }^{a}$, $a=1,...,N^{2}-N$, see \cite
{'tHooft:1981ht,Bruckmann:2000xd}. The gauge fixing condition for the
off-diagonal components $A_{\mu }^{a}$ can be obtained by requiring that the
functional $\mathcal{R}[A]$ is stationary under gauge transformations.
Moreover, a residual local $U(1)^{N-1}$ invariance, corresponding to the
Cartan subgroup of $SU(N)$, is still present \cite
{'tHooft:1981ht,Bruckmann:2000xd}. This local invariance has to be fixed by
imposing an additional condition on the diagonal components $A_{\mu }^{i}$
of the gauge field, which will be chosen to be of the Landau type, \textit{%
i.e.} $\partial _{\mu }A_{\mu }^{i}=0$. Analogously to the Landau and
Coulomb gauges, a complete gauge fixing would require the implementation of
the restriction of the domain of integration in the path integral to the
fundamental modular region for the maximal Abelian gauge, a task which is
beyond our present capabilities. As already underlined, we shall limit
ourselves to the restriction to the Gribov region, which turns out to
correspond to field configurations which are relative minima of $\mathcal{R}%
[A]$. \newline
\newline
The output of our results can be summarized as follows. The diagonal
component of the gluon propagator is found to display the characteristic
Gribov type behavior
\begin{equation}
\left\langle A_{\mu }(k)A_{\nu }(-k)\right\rangle =\frac{k^{2}}{k^{4}+\gamma
^{4}}\left( \delta _{\mu \nu }-\frac{k_{\mu }k_{\nu }}{k^{2}}\right) \;,
\label{gl}
\end{equation}
where $\gamma $ is the Gribov parameter and $A_{\mu }$ stands for the
diagonal component of the gauge field in the case of $SU(2)$,\textit{\ i.e.}
$A_{\mu }=$ $A_{\mu }^{3}$. The off-diagonal propagator turns out to be of
the Yukawa type, being given by
\begin{eqnarray}
\left\langle A_{\mu }^{a}(k)A_{\nu }^{b}(-k)\right\rangle &=&\delta ^{ab}%
\frac{1}{k^{2}+m^{2}}\left( \delta _{\mu \nu }-\frac{k_{\mu }k_{\nu }}{k^{2}}%
\right) \;,\;\;\;\;\;\;\;\;\;  \label{offg} \\
a,b &=&1,2\;.
\end{eqnarray}
where $m$ denotes the off-diagonal dynamical mass originating from the
dimension two condensate $\left\langle A_{\mu }^{a}A_{\mu }^{a}\right\rangle
$. One observes that both propagators are suppressed in the infrared. In the
case of the ghost propagator, we find that the off-diagonal component
exhibits infrared enhancement, namely
\begin{eqnarray}
\left. \mathcal{G}\left( k\right) \right| _{k=0} &\approx &\frac{\gamma ^{2}%
}{k^{4}}\;,  \label{offghg} \\
\mathcal{G}\left( k\right) &=&\frac{1}{2}\sum_{a}\left\langle \bar{c}%
^{a}(k)c^{a}(-k)\right\rangle \;,  \nonumber
\end{eqnarray}
where $(\bar{c}^{a},c^{a})\;$stand for the off-diagonal Faddeev-Popov
ghosts, see Appendix \ref{B}. Finally, the diagonal component of the ghost
propagator turns out to be not affected by the restriction to the first
horizon.

\section{The gauge fixing condition for the maximal Abelian gauge}

In order to discuss the gauge fixing condition let us first remind some
basic properties of the maximal Abelian gauge in the case of $SU(2)$. The
gauge field is decomposed into off-diagonal and diagonal components,
according to
\begin{equation}
\mathcal{A}_{\mu }=A_{\mu }^{a}T^{a}+A_{\mu }T^{3}\;,  \label{conn}
\end{equation}
where $T^{a}$, $a=1,2$, denote the off-diagonal generators of $SU(2)$, while
$T^{3}$ stands for the diagonal generator,
\begin{eqnarray}
\left[ T^{a},T^{b}\right] ~ &=&i~\varepsilon ^{ab}T^{3},  \nonumber \\
\left[ T^{3},T^{a}\right] ~ &=&i~\varepsilon ^{ab}T^{b},  \label{la1}
\end{eqnarray}
where
\begin{eqnarray}
\varepsilon ^{ab} &=&\varepsilon ^{ab3}\;,  \nonumber \\
\varepsilon ^{ac}\varepsilon ^{ad} &=&\delta ^{cd}\;.  \label{la2}
\end{eqnarray}
Similarly, for the field strength one has
\begin{equation}
\mathcal{F}_{\mu \nu }=F_{\mu \nu }^{a}T^{a}+F_{\mu \nu }T^{3}\;,  \label{fs}
\end{equation}
with the off-diagonal and diagonal parts given, respectively, by
\begin{eqnarray}
F_{\mu \nu }^{a} &=&D_{\mu }^{ab}A_{\nu }^{b}-D_{\nu }^{ab}A_{\mu }^{b}\;,
\label{fsc} \\
F_{\mu \nu } &=&\partial _{\mu }A_{\nu }-\partial _{\nu }A_{\mu
}+g\varepsilon ^{ab}A_{\mu }^{a}A_{\nu }^{b}\;,  \nonumber
\end{eqnarray}
where the covariant derivative $D_{\mu }^{ab}$ is defined with respect to
the diagonal component $A_{\mu }$
\begin{equation}
D_{\mu }^{ab}\equiv \partial _{\mu }\delta ^{ab}-g\varepsilon ^{ab}A_{\mu
}\;.  \label{cv}
\end{equation}
Thus, for the Yang-Mills action in Euclidean space one obtains
\begin{equation}
S_{\mathrm{YM}}=\frac{1}{4}\int d^{4}x\,\left( F_{\mu \nu }^{a}F_{\mu \nu
}^{a}+F_{\mu \nu }F_{\mu \nu }\right) \;.  \label{ym}
\end{equation}
As it is easily checked, the classical action (\ref{ym}) is left invariant
by the gauge transformations
\begin{eqnarray}
\delta A_{\mu }^{a} &=&-D_{\mu }^{ab}{\omega }^{b}-g\varepsilon ^{ab}A_{\mu
}^{b}\omega \;,  \nonumber \\
\delta A_{\mu } &=&-\partial _{\mu }{\omega }-g\varepsilon ^{ab}A_{\mu
}^{a}\omega ^{b}\;.  \label{gauge}
\end{eqnarray}
The maximal Abelian gauge is obtained by demanding that the off-diagonal
components $A_{\mu }^{a}$ of the gauge field obey the nonlinear condition
\begin{equation}
D_{\mu }^{ab}A_{\mu }^{b}=0\;,  \label{offgauge}
\end{equation}
which follows by requiring that the auxiliary functional
\begin{equation}
\mathcal{R}[A]=\int {d^{4}x}A_{\mu }^{a}A_{\mu }^{a}\;,  \label{fmag}
\end{equation}
is stationary with respect to the gauge transformations (\ref{gauge}).
Moreover, as it is apparent from the presence of the covariant derivative $%
D_{\mu }^{ab}$, equation (\ref{offgauge}) allows for a residual local $U(1)$
invariance corresponding to the diagonal subgroup of $SU(2)$ \cite
{Bruckmann:2000xd}. This additional invariance has to be fixed by means of a
suitable gauge condition on the diagonal component $A_{\mu }$, which will be
chosen to be of the Landau type, also adopted in lattice simulations, namely
\begin{equation}
\partial _{\mu }A_{\mu }=0\;.  \label{dgauge}
\end{equation}
Let us work out the condition for the existence of Gribov copies in the
maximal Abelian gauge. In the case of small gauge transformations, this is
easily obtained by requiring that the transformed fields, eqs.(\ref{gauge}),
fulfill the same gauge conditions obeyed by $\left( A_{\mu },A_{\mu
}^{a}\right) $, \textit{i.e.} eqs.(\ref{offgauge}), (\ref{dgauge}). Thus, to
the first order in the gauge parameters $\left( \omega ,\omega ^{a}\right) $%
, one gets
\begin{eqnarray}
-D_{\mu }^{ab}D_{\mu }^{bc}\omega ^{c}-g\varepsilon ^{bc}D_{\mu }^{ab}\left(
A_{\mu }^{c}\omega \right) +g\varepsilon ^{ab}A_{\mu }^{b}\partial _{\mu
}\omega +g^{2}\varepsilon ^{ab}\varepsilon ^{cd}A_{\mu }^{b}A_{\mu
}^{c}\omega ^{d} &=&0\;,  \label{offcopies0} \\
-\partial ^{2}\omega -g\varepsilon ^{ab}\partial _{\mu }\left( A_{\mu
}^{a}\omega ^{b}\right)  &=&0\;,  \label{diagcopies0}
\end{eqnarray}
which, due to eqs.(\ref{offgauge}),(\ref{dgauge}) read
\begin{eqnarray}
\mathcal{M}^{ab}\omega ^{b} &=&0\;,  \label{off1} \\
-\partial ^{2}\omega -g\varepsilon ^{ab}\partial _{\mu }\left( A_{\mu
}^{a}\omega ^{b}\right)  &=&0\;,  \label{de}
\end{eqnarray}
with $\mathcal{M}^{ab}$ given by
\begin{equation}
\mathcal{M}^{ab}=-D_{\mu }^{ac}D_{\mu }^{cb}-g^{2}\varepsilon
^{ac}\varepsilon ^{bd}A_{\mu }^{c}A_{\mu }^{d}\;.  \label{offop}
\end{equation}
The operator $\mathcal{M}^{ab}$ is recognized to be the Faddeev-Popov
operator \cite{Quandt:1997rg} for the off-diagonal ghost sector, see
Appendix \ref{B}. It enjoys the property of being Hermitian and, as pointed
out in \cite{Bruckmann:2000xd}, is the difference of two positive
semidefinite operators given, respectively, by $-D_{\mu }^{ac}D_{\mu }^{cb}$
and $g^{2}\varepsilon ^{ac}\varepsilon ^{bd}A_{\mu }^{c}A_{\mu }^{d}$. Also,
one should remark that the diagonal parameter $\omega $ appears only in the
eq.(\ref{de}), in a form which allows us to express it in terms of the
solution of the first equation (\ref{off1}). More precisely, once eq.(\ref
{off1}) has been solved for $A_{\mu }$, $A_{\mu }^{a}$, $\omega ^{b}$, for
the diagonal parameter $\omega $ one can write
\begin{equation}
\omega =-g\epsilon ^{ab}\frac{\partial _{\mu }}{\partial ^{2}}\left( A_{\mu
}^{a}\omega ^{b}\right) \;.  \label{des}
\end{equation}
This feature means essentially that the diagonal parameter $\omega $ has no
special role in the characterization of the Gribov copies, whose properties
are encoded in eq.(\ref{off1}). Also, from eq.(\ref{des}) it follows that
the new variable $\tilde{\omega}$%
\begin{equation}
\tilde{\omega}=\omega +g\epsilon ^{ab}\frac{\partial _{\mu }}{\partial ^{2}}%
\left( A_{\mu }^{a}\omega ^{b}\right) \;,  \label{trad}
\end{equation}
obeys
\begin{equation}
\partial ^{2}\tilde{\omega}=0\;.  \label{ab1}
\end{equation}
As shown in Appendix B, the change of variable (\ref{trad}) can be performed
in the partition function expressing the Faddeev-Popov quantization of
Yang-Mills theories in the maximal Abelian gauge. As the corresponding
Jacobian turns out to be independent from the fields, transformation (\ref
{trad}) has the effect of decoupling the diagonal ghost fields from the
theory. As a consequence, the corresponding two point function is not
affected by the restriction to the Gribov region.

\section{Restriction of the domain of integration to the Gribov region}

Let us face now the implementation in the Feynman path integral of the
restriction of the domain of integration to the Gribov region $\mathcal{C}%
_{0}$, defined as the set of fields fulfilling the gauge conditions (\ref
{offgauge}), (\ref{dgauge}) and for which the Faddeev-Popov operator $%
\mathcal{M}^{ab}$ is positive definite, namely
\begin{equation}
\mathcal{C}_{0}=\left\{ A_{\mu },\;A_{\mu }^{a},\;\partial _{\mu }A_{\mu
}=0,\;D_{\mu }^{ab}A_{\mu }^{b}=0,\;\mathcal{M}^{ab}=-D_{\mu }^{ac}D_{\mu
}^{cb}-g^{2}\varepsilon ^{ac}\varepsilon ^{bd}A_{\mu }^{c}A_{\mu
}^{d}>0\right\} \;.  \label{gr}
\end{equation}
The boundary, $l_{1}$, of the region $\mathcal{C}_{0}$, where the first
vanishing eigenvalue of $\mathcal{M}^{ab}$ appears, is called the first
Gribov horizon. The restriction of the domain of integration to this region
is supported by the possibility of generalizing to the maximal Abelian gauge
Gribov's original result \cite{Gribov:1977wm} stating that for any field
located near a horizon there is a gauge copy, close to the same horizon,
located on the other side of the horizon. We have found useful to devote the
whole Appendix \ref{A} to the details of the proof of this statement.
\newline
\newline
Thus, for the partition function of Yang-Mills theory in the maximal Abelian
gauge, we write
\begin{equation}
\mathcal{Z}=\int DA_{\mu }^{a}DA_{\mu }\;\det \left( \mathcal{M}%
^{ab}(A)\right) \;\delta \left( D_{\mu }^{ab}A_{\mu }^{b}\right) \delta
\left( \partial _{\mu }A_{\mu }\right) e^{-S_{YM}}\mathcal{V}(\mathcal{C}%
_{0})\;,  \label{pf}
\end{equation}
where the factor $\mathcal{V}(\mathcal{C}_{0})$ implements the restriction
to the region $\mathcal{C}_{0}$. Following \cite{Gribov:1977wm}, the factor $%
\mathcal{V}(\mathcal{C}_{0})$ can be accommodated for by means of a no pole
condition on the off-diagonal ghost two-point function, given by the inverse
of the Faddeev-Popov operator $\mathcal{M}^{ab}$. More precisely, denoting
by $\mathcal{G}(k,A)$ the Fourier transform of $\left( \mathcal{M}%
^{ab}\right) ^{-1}$, \textit{i.e.}
\begin{equation}
\mathcal{G}\left( k,A\right) =\frac{1}{2}\sum_{ab}\delta ^{ab}\left\langle
k\left| \left( \mathcal{M}^{ab}\right) ^{-1}\right| k\right\rangle \;,
\label{gh2}
\end{equation}
we shall require that $\mathcal{G}\left( k,A\right) $ has no poles for a
given nonvanishing value of the momentum $k$, except for a singularity at $%
k=0$, corresponding to the boundary of $\mathcal{C}_{0}$, \textit{i.e. }to
the first Gribov horizon $l_{1}$ \cite{Gribov:1977wm}. This no pole
condition can be easily understood by observing that, within the region $%
\mathcal{C}_{0}$, the Faddeev-Popov operator $\mathcal{M}^{ab}$ is positive
definite. This implies that its inverse, $\left( \mathcal{M}^{ab}\right)
^{-1}$, and thus the Green function $\mathcal{G}$ of eq.$\left( \ref{gh2}%
\right) $, can become large only when approaching the horizon $l_{1}$, where
the operator $\mathcal{M}^{ab}$ has a zero mode. \newline
\newline
The Green function $\mathcal{G}$ can be evaluated order by order. Repeating
the same procedure of \cite{Gribov:1977wm} in the case of the maximal
Abelian gauge, we find that, up to the second order,
\begin{equation}
\mathcal{G}\left( k,A\right) =\frac{1}{k^{2}}+g^{2}\frac{k_{\mu }k_{\nu }}{%
k^{4}}\frac{1}{V}\sum_{q}\frac{A_{\mu }(q)A_{\nu }(-q)}{\left( k-q\right)
^{2}}+\frac{g^{2}}{k^{4}}\frac{1}{V}\sum_{q}A_{\mu }(q)A_{\mu }(-q)+\frac{%
g^{2}}{2k^{4}}\frac{1}{V}\sum_{q}A_{\mu }^{a}(q)A_{\mu }^{a}(-q)\;
\label{seco}
\end{equation}
where $V$ is the Euclidean volume. We observe that the last two terms of
expression $\left( \ref{seco}\right) $, \textit{i.e. }$\sum_{q}A_{\mu
}(q)A_{\mu }(-q)$ and $\sum_{q}A_{\mu }^{a}(q)A_{\mu }^{a}(-q)$, do not
depend on the external momentum $k$. Therefore, after subtraction of the
corresponding ultraviolet perturbative parts\footnote{%
At the perturbative level, these terms give rise to tadpole contributions.
As such, they vanish in dimensional regularization, which will be implicitly
employed throughout.}, these terms might yield a nonperturbative
contribution to the Green function $\mathcal{G}$, corresponding to the
singularity at $k=0$, as it is apparent from the presence of the factor $%
1/k^{4}$ in eq.$\left( \ref{seco}\right) $. We shall see in fact that these
terms will give rise to a nonperturbative contribution which is proportional
to the Gribov parameter $\gamma $. \newline
\newline
Thus, for $\mathcal{G}\left( k,A\right) $ we shall write \cite{Gribov:1977wm}
\begin{equation}
\mathcal{G}\left( k,A\right) \approx \frac{1}{k^{2}}\frac{1}{\left[ 1-\sigma
\left( k,A\right) \right] }+\frac{\mathcal{B}}{k^{4}}\;,  \label{s1}
\end{equation}
where
\begin{eqnarray}
\sigma \left( k,A\right) &=&\frac{g^{2}}{V}\frac{k_{\mu }k_{\nu }}{k^{2}}%
\sum_{q}\frac{A_{\mu }(q)A_{\nu }(-q)}{\left( k-q\right) ^{2}}\;,  \nonumber
\\
\mathcal{B} &=&\frac{g^{2}}{V}\sum_{q}A_{\mu }(q)A_{\mu }(-q)+\frac{g^{2}}{V}%
\sum_{q}A_{\mu }^{a}(q)A_{\mu }^{a}(-q)\;,  \label{s2}
\end{eqnarray}
which, in the thermodynamic limit, $V\rightarrow \infty $, become
\begin{eqnarray}
\sigma \left( k,A\right) &=&g^{2}\frac{k_{\mu }k_{\nu }}{k^{2}}\int \frac{%
d^{4}q}{\left( 2\pi \right) ^{4}}\frac{A_{\mu }(q)A_{\nu }(-q)}{\left(
k-q\right) ^{2}}\;,  \nonumber \\
\mathcal{B} &=&g^{2}\int \frac{d^{4}q}{\left( 2\pi \right) ^{4}}A_{\mu
}(q)A_{\mu }(-q)+\frac{g^{2}}{2}\int \frac{d^{4}q}{\left( 2\pi \right) ^{4}}%
A_{\mu }^{a}(q)A_{\mu }^{a}(-q)\;.  \label{s3}
\end{eqnarray}
The expression for $\sigma \left( k,A\right) $ in eq.$\left( \ref{s3}\right)
$ can be simplified by recalling that, due to the Landau gauge condition,
the Abelian component $A_{\mu }(q)$ is transverse, namely
\begin{equation}
q_{\mu }A_{\mu }(q)=0\;.  \label{tr1}
\end{equation}
Setting
\begin{eqnarray}
A_{\mu }(q)A_{\nu }(-q) &=&\omega (A)\left( \delta _{\mu \nu }-\frac{q_{\mu
}q_{\nu }}{q^{2}}\right) \;,  \nonumber \\
\omega (A) &=&\frac{1}{3}A_{\lambda }(q)A_{\lambda }(-q)\;,  \label{tr2}
\end{eqnarray}
for $\sigma \left( k,A\right) $ one obtains
\begin{equation}
\sigma \left( k,A\right) \text{ }=g^{2}\frac{k_{\mu }k_{\nu }}{k^{2}}\frac{1%
}{3}\int \frac{d^{4}q}{\left( 2\pi \right) ^{4}}\frac{A_{\lambda
}(q)A_{\lambda }(-q)}{\left( k-q\right) ^{2}}\left( \delta _{\mu \nu }-\frac{%
q_{\mu }q_{\nu }}{q^{2}}\right) \;.  \label{sg}
\end{equation}
Note that expression $\left( \ref{sg}\right) $ is, in practice, the same as
that obtained by Gribov \cite{Gribov:1977wm} in the case of the Landau
gauge. This is not surprising since $\sigma \left( k,A\right) $ depends only
on the diagonal component $A_{\mu }(q)$, which is in fact transverse.
Finally, following \cite{Gribov:1977wm}, the no-pole condition at finite
nonvanishing $k$ for the Green function $\mathcal{G}\left( k,A\right) $ can
be stated as
\begin{equation}
\sigma \left( 0,A\right) <1\;,  \label{np}
\end{equation}
with
\begin{equation}
\sigma \left( 0,A\right) =g^{2}\frac{1}{4}\int \frac{d^{4}q}{\left( 2\pi
\right) ^{4}}\frac{A_{\lambda }(q)A_{\lambda }(-q)}{q^{2}}\;,  \label{nps}
\end{equation}
where use has been made of
\begin{equation}
\int \frac{d^{4}q}{\left( 2\pi \right) ^{4}}\frac{A_{\lambda }(q)A_{\lambda
}(-q)}{q^{2}}\left( \delta _{\mu \nu }-\frac{q_{\mu }q_{\nu }}{q^{2}}\right)
=\frac{3}{4}\delta _{\mu \nu }\int \frac{d^{4}q}{\left( 2\pi \right) ^{4}}%
\frac{A_{\lambda }(q)A_{\lambda }(-q)}{q^{2}}\;,  \label{ltz}
\end{equation}
which follows from Lorentz covariance. Condition $\left( \ref{np}\right) $
ensures that the Green function $\mathcal{G}\left( k,A\right) $ in eq.$%
\left( \ref{s1}\right) $ has no poles at finite nonvanishing $k$. The only
allowed singularity is that at $k=0$, corresponding to approaching the first
Gribov horizon $l_{1}$.

\subsection{The gluon propagator}

We are now ready to discuss the behavior of the gluon propagator when the
domain of integration in the Feynman path integral is restricted to the
region $\mathcal{C}_{0}$, eq.$\left( \ref{pf}\right) $. According to \cite
{Gribov:1977wm}, the factor $\mathcal{V}(\mathcal{C}_{0})$ implementing the
restriction to $\mathcal{C}_{0}$ is given by
\begin{equation}
\mathcal{V}(\mathcal{C}_{0})=\theta \left[ 1-\sigma (0,A)\right] \;,
\label{theta0}
\end{equation}
where $\theta (x)$ stands for the step function\footnote{$\theta (x)=1$ for $%
x>0$, and $\theta (x)=0$ for $x<0$.}. Moreover, making use of the integral
representation
\begin{equation}
\theta (1-\sigma (0,A))=\int_{-i\infty +\varepsilon }^{i\infty +\varepsilon }%
\frac{d\zeta }{2\pi i\zeta }e^{\zeta (1-\sigma (0,A))}\;,  \label{p1}
\end{equation}
for the partition function $\mathcal{Z}$ we get
\begin{equation}
\mathcal{Z=}\int DA_{\mu }^{a}DA_{\mu }\frac{d\zeta }{2\pi i\zeta }\det
\left( \mathcal{M}^{ab}(A)\right) \exp \left( \zeta -S_{\mathrm{YM}}-\frac{1%
}{2\alpha }\left( D_{\mu }^{ab}A_{\mu }^{b}\right) ^{2}-\frac{1}{2\beta }%
\left( \partial _{\mu }A_{\mu }\right) ^{2}-\zeta \sigma (0,A)\right) \;
\label{pf1}
\end{equation}
where the gauge parameters $\alpha $ and $\beta $ have to be set to zero at
end, \textit{i.e. }$\alpha $, $\beta \rightarrow 0$, to recover the gauge
conditions (\ref{offgauge}), (\ref{dgauge}). In order to study the gluon
propagator, it is sufficient to retain only the quadratic terms in
expression $\left( \ref{pf1}\right) $ which contribute to the two-point
correlation functions $\left\langle A_{\mu }^{a}(k)A_{\nu
}^{b}(-k)\right\rangle $ and $\left\langle A_{\mu }(k)A_{\nu
}(-k)\right\rangle $. Thus
\begin{equation}
\mathcal{Z}_{\mathrm{quadr}}=\mathcal{N}\int DA_{\mu }^{a}DA_{\mu }\frac{%
d\zeta }{2\pi i}e^{\left( \zeta -\log \zeta -S_{\mathrm{quadr}}-\zeta \sigma
(0,A)\right) }\;,  \label{pfq}
\end{equation}
where $\mathcal{N}$ is a constant factor and $S_{\mathrm{quadr}}$ stands for
the quadratic part of the quantized Yang-Mills action, namely
\begin{eqnarray}
S_{\mathrm{quadr}} &=&\frac{1}{2}\sum_{q}\left( A_{\mu }^{a}(q)\left(
q^{2}\delta _{\mu \nu }-\left( 1-\frac{1}{\alpha }\right) q_{\mu }q_{\nu
}\right) A_{\nu }^{a}(-q)\right)  \nonumber \\
&&+\;\frac{1}{2}\sum_{q}\left( A_{\mu }(q)\left( q^{2}\delta _{\mu \nu
}-\left( 1-\frac{1}{\beta }\right) q_{\mu }q_{\nu }\right) A_{\nu
}(-q)\right) \;.  \label{qp}
\end{eqnarray}
Therefore, recalling the expression for the factor $\sigma (0,A)$, eq.(\ref
{nps}), it follows
\begin{equation}
\mathcal{Z}_{\mathrm{quadr}}=\mathcal{N}\int DA_{\mu }^{a}DA_{\mu }\frac{%
d\zeta }{2\pi i}\exp \left( {\zeta -\log \zeta -}\frac{1}{2}\sum_{q}A{_{\mu
}(}q{)\mathcal{Q}_{\mu \nu }(\zeta ,}q{)}A{_{\nu }(-}q{)-}\;\frac{1}{2}%
\sum_{q}A_{\mu }^{a}(q{)\mathcal{P}_{\mu \nu }(}q{)}A_{\mu }^{a}(-q{)}%
\right) \;  \label{qp2}
\end{equation}
where the quantities ${\mathcal{Q}_{\mu \nu }(\zeta ,}q{)}$ and ${\mathcal{P}%
_{\mu \nu }(}q{)}$ are given by
\begin{eqnarray}
\mathcal{Q}_{\mu \nu }(\zeta ,q) &=&\left( q^{2}+\frac{\zeta g^{2}}{2Vq^{2}}%
\right) \delta _{\mu \nu }-\left( 1-\frac{1}{\beta }\right) q_{\mu }q_{\nu
}\;,  \nonumber \\
\mathcal{P}_{\mu \nu }(q) &=&q^{2}\delta _{\mu \nu }-\left( 1-\frac{1}{%
\alpha }\right) q_{\mu }q_{\nu }\;.  \label{fqp}
\end{eqnarray}
Note that only the factor $\mathcal{Q}_{\mu \nu }$, corresponding to the
operator appearing in the quadratic part for the diagonal component $A{_{\mu
}(}q)$ in eq.(\ref{qp2}), depends on $\zeta $. Integrating over the gauge
fields and keeping only the terms which depend on $\zeta $, we find
\begin{equation}
\mathcal{Z}_{\mathrm{quadr}}=\mathcal{N}\int \frac{d\zeta }{2\pi i}e^{\zeta
-\log \zeta }\left( \det \mathcal{Q}_{\mu \nu }(\zeta ,q)\right)
^{-1/2}\left( \det \mathcal{P}_{\mu \nu }(q)\right) ^{-1}=\mathcal{N}%
^{\prime }\int \frac{d\zeta }{2\pi i}e^{f(\zeta )}\;,  \label{qpi}
\end{equation}
where
\begin{eqnarray}
f(\zeta ) &=&\zeta -\log \zeta -\frac{1}{2}\log \det \left( \mathcal{Q}_{\mu
\nu }(\zeta ,q)\;\right) ,  \nonumber \\
&=&\zeta -\log \zeta -\frac{3}{2}\sum_{q}\log \left( q^{2}+\frac{\zeta g^{2}%
}{2Vq^{2}}\right) \;.  \label{fcond}
\end{eqnarray}
As done in \cite{Gribov:1977wm}, expression $\left( \mathrm{{\ref{qpi}}}%
\right) $ can be now evaluated at the saddle point, namely
\begin{equation}
\mathcal{Z}_{\mathrm{quadr}}\approx e^{f(\zeta _{0})}\;,  \label{sp}
\end{equation}
where $\zeta _{0}$ is determined by the minimum condition
\begin{equation}
\left. \frac{\partial f(\zeta )}{\partial \zeta }\right| _{\zeta =\zeta
_{0}}=0\;,  \label{mc}
\end{equation}
which yields
\begin{equation}
1-\frac{1}{\zeta _{0}}-\frac{3g^{2}}{4V}\sum_{q}\frac{1}{q^{4}+\frac{\zeta
_{0}g^{2}}{2V}}=0\;.  \label{mcr}
\end{equation}
Taking the thermodynamic limit, $V\rightarrow \infty $, and introducing the
Gribov parameter $\gamma $ \cite{Gribov:1977wm}
\begin{equation}
\gamma ^{4}=\frac{\zeta _{0}g^{2}}{2V}\;,\;\;\;\;\;\;V\rightarrow \infty \;,
\label{gbg}
\end{equation}
we get the gap equation
\begin{equation}
\frac{3}{4}g^{2}\int \frac{d^{4}q}{\left( 2\pi \right) ^{4}}\frac{1}{%
q^{4}+\gamma ^{4}}=1\;,  \label{gapmag}
\end{equation}
where the term $1/\zeta _{0}$ in eq.$\left( \mathrm{{\ref{mcr}}}\right) $
has been neglected in the thermodynamic limit. To obtain the gauge
propagator, we can now go back to the expression for $\mathcal{Z}_{\mathrm{%
quadr}}$ which, after substituting the saddle point value $\zeta =\zeta _{0}$%
, becomes
\begin{equation}
\mathcal{Z}_{\mathrm{quadr}}=\mathcal{N}\int DA_{\mu }^{a}DA_{\mu }e^{-\frac{%
1}{2}\left( \sum_{q}A{_{\mu }(}q{)\mathcal{Q}_{\mu \nu }(\gamma ,}q{)}A{%
_{\nu }(-}q{)+}\sum_{q}A_{\mu }^{a}(q{)\mathcal{P}_{\mu \nu }(}q{)}A_{\mu
}^{a}(-q)\right) \;},  \label{zqf}
\end{equation}
with
\begin{equation}
\mathcal{Q}_{\mu \nu }(\gamma ,q)=\left( q^{2}+\frac{\gamma ^{4}}{q^{2}}%
\right) \delta _{\mu \nu }-\left( 1-\frac{1}{\beta }\right) q_{\mu}q_{\nu
}\;.  \label{qg}
\end{equation}
Evaluating the inverse of $\mathcal{Q}_{\mu \nu }(\gamma ,q)$ and of ${%
\mathcal{P}_{\mu \nu }(}q{)}$, and setting the gauge parameters $\alpha $,$%
\beta $ to zero, we get the gluon propagator for the diagonal and off
diagonal components of the gauge field, namely
\begin{equation}
\left\langle A_{\mu }(q)A_{\nu }(-q)\right\rangle =\frac{q^{2}}{q^{4}+\gamma
^{4}}\left( \delta _{\mu \nu }-\frac{q_{\mu }q_{\nu }}{q^{2}}\right) \;,
\label{dprop}
\end{equation}
and
\begin{equation}
\left\langle A_{\mu }^{a}(q)A_{\nu }^{b}(-q)\right\rangle =\delta ^{ab}\frac{%
1}{q^{2}}\left( \delta _{\mu \nu }-\frac{q_{\mu }q_{\nu }}{q^{2}}\right) \;,
\label{offdp}
\end{equation}
One sees that the diagonal component, eq.$\left( \mathrm{{\ref{dprop}}}%
\right) $, is suppressed in the infrared, exhibiting the characteristic
Gribov type behavior. The off-diagonal components, eq.$\left( \mathrm{{\ref
{offdp}}}\right) $, remains unchanged. Moreover, as we shall see later, its
infrared behavior turns out to be modified once the gluon condensate $%
\left\langle A_{\mu }^{a}A_{\mu }^{a}\right\rangle $ is taken into account.
\newline

\subsection{The off-diagonal ghost propagator}

The off-diagonal ghost propagator can be obtained from eq.$\left( \ref{s1}%
\right) $ upon contraction of the gauge fields in expressions $\left( \ref
{s3}\right) $, namely
\begin{equation}
\mathcal{G}\left( k\right) \approx \frac{1}{k^{2}}\frac{1}{\left[ 1-\sigma
\left( k\right) \right] }+\frac{\mathcal{B}}{k^{4}}\;,  \label{ghp}
\end{equation}
with
\begin{equation}
\sigma \left( k\right) =g^{2}\frac{k_{\mu }k_{\nu }}{k^{2}}\int \frac{d^{4}q%
}{\left( 2\pi \right) ^{4}}\frac{\left\langle A_{\mu }(q)A_{\nu
}(-q)\right\rangle }{\left( k-q\right) ^{2}}\;,  \label{cs1}
\end{equation}
and
\begin{equation}
\mathcal{B}=g^{2}\int \frac{d^{4}q}{\left( 2\pi \right) ^{4}}\left\langle
A_{\mu }(q)A_{\mu }(-q)\right\rangle +\frac{g^{2}}{2}\int \frac{d^{4}q}{%
\left( 2\pi \right) ^{4}}\left\langle A_{\mu }^{a}(q)A_{\mu
}^{a}(-q)\right\rangle \;.  \label{cs2}
\end{equation}
Let us consider first the factor $\sigma \left( k\right) $ of eq.$\left( \ref
{cs1}\right) $. From the expression of the diagonal propagator in eq.$\left(
\ref{dprop}\right) $, we obtain

\begin{equation}
\sigma \left( k\right) =g^{2}\frac{k_{\mu }k_{\nu }}{k^{2}}\int \frac{d^{4}q%
}{\left( 2\pi \right) ^{4}}\frac{q^{2}}{\left( k-q\right) ^{2}\left(
q^{4}+\gamma ^{4}\right) }\left( \delta _{\mu \nu }-\frac{q_{\mu }q_{\nu }}{%
q^{2}}\right) \;.  \label{sigmak1}
\end{equation}
Making use of the gap equation (\ref{gapmag}), we can write
\begin{equation}
g^{2}\frac{k_{\mu }k_{\nu }}{k^{2}}\int \frac{d^{4}q}{\left( 2\pi \right)
^{4}}\frac{1}{q^{4}+\gamma ^{4}}\left( \delta _{\mu \nu }-\frac{q_{\mu
}q_{\nu }}{q^{2}}\right) =1\;,  \label{gap2}
\end{equation}
so that
\begin{equation}
1-\sigma \left( k\right) =g^{2}\frac{k_{\mu }k_{\nu }}{k^{2}}\int \frac{%
d^{4}q}{\left( 2\pi \right) ^{4}}\frac{k^{2}-2kq}{\left( k-q\right)
^{2}\left( q^{4}+\gamma ^{4}\right) }\left( \delta _{\mu \nu }-\frac{q_{\mu
}q_{\nu }}{q^{2}}\right) \;.  \label{bla0}
\end{equation}
Note that the integral in eq.(\ref{bla0}) is ultraviolet finite. Thus, in
the infrared, $k\approx 0$, one gets
\begin{equation}
\left. \left( 1-\sigma \left( k\right) \right) \right| _{k\approx 0}=\frac{%
3g^{2}k^{2}}{4}\int \frac{d^{4}q}{\left( 2\pi \right) ^{4}}\frac{1}{%
q^{2}\left( q^{4}+\gamma ^{4}\right) }=\frac{3g^{2}k^{2}}{128\pi \gamma ^{2}}%
\;.  \label{bla1}
\end{equation}
It remains now to discuss the factor $\mathcal{B}$ of eq.$\left( \ref{cs2}%
\right) $. Making use of the dimensional regularization in the $\overline{MS}
$ scheme, one observes that, due to the form of the off-diagonal propagator,
eq.$\left( \ref{offdp}\right) $, the second term of eq.$\left( \ref{cs2}%
\right) $ vanishes. Concerning now the first term, it is not difficult to
see that it gives a contribution proportional to the Gribov parameter $%
\gamma ^{2}$. In fact
\begin{equation}
\int \frac{d^{4}q}{\left( 2\pi \right) ^{4}}\left\langle A_{\mu }(q)A_{\mu
}(-q)\right\rangle =3\int \frac{d^{4}q}{\left( 2\pi \right) ^{4}}\frac{q^{2}%
}{q^{4}+\gamma ^{4}}=-3\gamma ^{4}\int \frac{d^{4}q}{\left( 2\pi \right) ^{4}%
}\frac{1}{q^{2}\left( q^{4}+\gamma ^{4}\right) }=-\frac{3\gamma ^{2}}{32\pi }%
\;.  \label{bt}
\end{equation}
Finally, for the infrared behavior of the off-diagonal ghost propagator we
have
\begin{equation}
\mathcal{G}(k)_{k\approx 0}\approx \left( \frac{128\pi }{3g^{2}}-\frac{3g^{2}%
}{32\pi }\right) \frac{\gamma ^{2}}{k^{4}}\;.  \label{offgh}
\end{equation}
exhibiting infrared enhancement.

\section{Inclusion of the dimension two condensate $\left\langle A_{\mu
}^{a}A_{\mu }^{a}\right\rangle $}

In this section we shall discuss the behavior of the propagators
when the dimension two condensate $\left\langle A_{\mu }^{a}A_{\mu
}^{a}\right\rangle $ is taken into account. This condensate turns
out to contribute to the gluon two-point function, as observed in
\cite{Lavelle:1988eg} within the operator product expansion. As
such, it has to be taken into account when discussing the gluon
propagator.  \newline
\newline
A renormalizable effective potential for $\left\langle A_{\mu }^{a}A_{\mu
}^{a}\right\rangle $ in the maximal Abelian gauge has been constructed and
evaluated in analytic form in \cite{Dudal:2004rx}. A nonvanishing condensate
$\left\langle A_{\mu }^{a}A_{\mu }^{a}\right\rangle $ is favoured since it
lowers the vacuum energy. As a consequence, a dynamical tree level mass for
off-diagonal gluons is generated. The inclusion of the condensate $%
\left\langle A_{\mu }^{a}A_{\mu }^{a}\right\rangle $ in the present
framework can be done along the lines outlined in \cite
{Sobreiro:2004us,Dudal:2005na}, where the effects of the Gribov copies on
the gluon and ghost propagators in the presence of the dimension two gluon
condensate have been worked out in the Landau gauge. Let us begin by giving
a brief account of the dynamical mass generation in the maximal Abelian
gauge. Following \cite{Dudal:2004rx}, the dynamical mass generation is
accounted for by adding to the gauge fixed Yang-Mills action the following
term
\begin{equation}
S_{\sigma }=\int d^{4}x\left( \frac{\sigma ^{2}}{2g^{2}\zeta }+\frac{1}{2}%
\frac{\sigma }{g\zeta }A_{\mu }^{a}A_{\mu }^{a}+\frac{1}{8\zeta }\left(
A_{\mu }^{a}A_{\mu }^{a}\right) ^{2}\;\right) .  \label{m2}
\end{equation}
The field $\sigma $ is an auxiliary field which allows one to study the
condensation of the local operator $A_{\mu }^{a}A_{\mu }^{a}$. In fact, as
shown in \cite{Dudal:2004rx}, the following relation holds
\begin{equation}
\left\langle \sigma \right\rangle =-\frac{g}{2}\left\langle A_{\mu
}^{a}A_{\mu }^{a}\right\rangle \;.  \label{m3}
\end{equation}
The dimensionless parameter $\zeta $ in expression $\left( \ref{m2}\right) $
is needed to account for the ultraviolet divergences present in the vacuum
correlation function $\left\langle A^{2}(x)A^{2}(y)\right\rangle $. For the
details of the renormalizability properties of the local operator $A_{\mu
}^{a}A_{\mu }^{a}$ in the maximal Abelian gauge we refer to \cite
{Dudal:2004rx,Gracey:2005vu,Kondo:2001tm,Dudal:2003pe}. The inclusion of the
term $S_{\sigma }$ is the starting point for evaluating the renormalizable
effective potential $V(\sigma )$ for the auxiliary field $\sigma $, obeying
the renormalization group equations. The minimum of $V(\sigma )$ occurs for
a nonvanishing vacuum expectation value of the auxiliary field, \textit{i.e.
}$\left\langle \sigma \right\rangle \neq 0$. In particular, the first order
off-diagonal dynamical gluon mass
\begin{equation}
m^{2}=\frac{\left\langle \sigma \right\rangle }{g\zeta }\;,  \label{msig}
\end{equation}
turns out to be \cite{Dudal:2004rx}
\begin{equation}
m=\left( \frac{3}{2}e^{\frac{17}{6}}\right) ^{\frac{1}{4}}\Lambda _{%
\overline{MS}}\approx 2.25\;\Lambda _{\overline{MS}}\;.  \label{mv}
\end{equation}
The inclusion of the action $S_{\sigma }$ leads to a partition function
which is still plagued by the Gribov copies. It might be useful to note in
fact that $S_{\sigma }$ is left invariant by the local gauge transformations
\begin{eqnarray}
\delta A_{\mu }^{a} &=&-D_{\mu }^{ab}{\omega }^{b}-g\varepsilon ^{ab}A_{\mu
}^{b}\omega \;,  \nonumber \\
\delta A_{\mu } &=&-\partial _{\mu }{\omega }-g\varepsilon ^{ab}A_{\mu
}^{a}\omega ^{b}\;,  \nonumber \\
\delta \sigma &=&gA_{\mu }^{a}D_{\mu }^{ab}{\omega }^{b}\;,  \label{str}
\end{eqnarray}
and
\begin{equation}
\delta S_{\sigma }=0\;.  \label{m7}
\end{equation}
Therefore, implementing the restriction to the region $\mathcal{C}_{0}$, for
the partition function we obtain now
\begin{equation}
\mathcal{Z}=\int DA_{\mu }^{a}DA_{\mu }\;\det \left( \mathcal{M}%
^{ab}(A)\right) \;\delta \left( D_{\mu }^{ab}A_{\mu }^{b}\right) \delta
\left( \partial _{\mu }A_{\mu }\right) e^{-\left( S_{YM}+S_{\sigma }\right) }%
\mathcal{V}(\mathcal{C}_{0})\;.  \label{pfm}
\end{equation}
To discuss the gluon propagator we proceed as before and retain only the
quadratic terms in expression $\left( \ref{pfm}\right) $ which contribute to
the two-point correlation functions. Expanding around the nonvanishing
vacuum expectation value of the auxiliary field, $\left\langle \sigma
\right\rangle \neq 0$, one easily gets

\begin{eqnarray}
\mathcal{Z}_{\mathrm{quadr}} &=&\mathcal{N}\int DA_{\mu }^{a}DA_{\mu }\frac{%
d\zeta }{2\pi i} e^{\left( {\zeta -\log \zeta -}\frac{1}{2}\sum_{q}A{_{\mu }(%
}q{)\mathcal{Q}_{\mu \nu }(\zeta ,}q{)}A{_{\nu }(-}q{)-}\;\frac{1}{2}%
\sum_{q}A_{\mu }^{a}(q{)}\mathcal{P}^{m}_{\mu \nu }(q)A_{\mu }^{a}(-q{)}%
\right) }\;  \nonumber \\
&&  \label{zqm}
\end{eqnarray}
where the factor ${\mathcal{Q}_{\mu \nu }(\zeta ,}q{)}$ is the same as given
in eq.$\left( \ref{fqp}\right) $, while

\begin{equation}
\mathcal{P}_{\mu \nu }^{m}(q)=q^{2}\delta _{\mu \nu }+m^{2}\delta _{\mu \nu
}-\left( 1-\frac{1}{\alpha }\right) q_{\mu }{\ }q_{\nu }\;.  \label{pm}
\end{equation}
One sees that the inclusion of the dynamical mass $m$, due to the gluon
condensate $\left\langle A_{\mu }^{a}A_{\mu }^{a}\right\rangle $, affects
only the off-diagonal sector. As a consequence, the gap equation defining
the Gribov parameter $\gamma $, eq.$\left( \ref{gapmag}\right) $, and the
diagonal gluon propagator, eq.$\left( \ref{dprop}\right) $, will be not
affected by the dynamical mass $m$, thus remaining the same. However, the
mass $m$ enters now the expression for the off-diagonal gluon propagator,
which becomes of the Yukawa type, as given in expression (\ref{offg}).
Note that, when the gluon condensate is taken into account, both diagonal
and off-diagonal components of the gluon propagator are suppressed in the
low momentum region. Finally, the infrared behavior of the ghost propagator
is easily seen to display infrared enhancement

\begin{equation}
\mathcal{G}(k)_{k\approx 0}\approx 1/k^{4}.  \label{gha2}
\end{equation}

\section{Comparison with lattice numerical simulations}

Having discussed the infrared behavior of the gluon and ghost propagators,
as expressed by eqs.$\left( \ref{gl}\right) $,$\left( \ref{offg}\right) $
and by eq.$\left( \ref{offgh}\right) $, it is worth making a comparison with
the results available from numerical lattice simulations. \newline
\newline
The first study of the gluon propagator on the lattice in the maximal
Abelian gauge was made in \cite{Amemiya:1998jz}, in the case of $SU(2)$. The
gluon propagator was analysed in coordinate space and the Landau gauge was
employed in the diagonal sector. The off-diagonal component of the gluon
propagator was found to be short-ranged, exhibiting a Yukawa type behavior,
\textit{i.e.} displaying an exponentially suppression at large distances by
an effective mass $m_{off}\approx 1.2\;GeV$. The diagonal component of the
gluon propagator was found to propagate over larger distances, see Fig.1 and
Fig.2 of \cite{Amemiya:1998jz}. These results were interpreted as evidence
for the infrared Abelian dominance \cite
{Ezawa:bf,Suzuki:1989gp,Suzuki:1992gz,Hioki:1991ai}, supporting the dual
superconductivity picture for color confinement. \newline
\newline
More recently, a numerical investigation of the gluon propagator in the
maximal Abelian gauge has been worked out in \cite{Bornyakov:2003ee}. Also
here, the gauge group is $SU(2)$ and the Landau gauge has been used for the
diagonal sector. Moreover, the gluon propagator has been investigated now in
momentum space, a feature which allows for a more direct comparison with our
findings. The results obtained in \cite{Bornyakov:2003ee} show that, at low
momenta, the diagonal component of the gluon propagator is much larger than
the off-diagonal one. Several possible fits were studied for the components
of the gluon propagator. In particular, among the two parameter fits
proposed in \cite{Bornyakov:2003ee}, a Gribov like fit, see eq.(20) of \cite
{Bornyakov:2003ee}, \textit{i.e.}
\begin{equation}
D_{diag}(q)=\frac{Z_{dg}\;q^{2}}{q^{4}+m_{dg}^{4}}\;,  \label{gfit}
\end{equation}
turns out to be suitable for the diagonal component of the gluon propagator.
For off-diagonal gluons, a Yukawa type fit, see eq.(18) of \cite
{Bornyakov:2003ee}, \textit{i.e.}
\begin{equation}
D_{off}(q)=\frac{Z_{off}}{q^{2}+m_{off}^{2}}\;,  \label{yfit}
\end{equation}
seems to be well succeeded. The scalar functions, $D_{diag}$ and $D_{off}$,
in eqs.$\left( \ref{gfit}\right) $, $\left( \ref{yfit}\right) $ parametrize
the diagonal and off-diagonal transverse components of the gluon propagator
in the low momentum region
\begin{eqnarray}
\left\langle A_{\mu }(q)A_{\nu }(-q)\right\rangle &=&D_{diag}(q)\left(
\delta _{\mu \nu }-\frac{q_{\mu }q_{\nu }}{q^{2}}\right) \;,  \nonumber \\
\left\langle A_{\mu }^{a}(q)A_{\nu }^{b}(-q)\right\rangle &=&\delta
^{ab}D_{off}(q)\left( \delta _{\mu \nu }-\frac{q_{\mu }q_{\nu }}{q^{2}}%
\right) \;.  \label{param}
\end{eqnarray}
The mass parameter $m_{off}$ appearing in the Yukawa fit $\left( \ref{yfit}%
\right) $ is two times bigger that the corresponding mass parameter $m_{dg}$
of the Gribov fit $\left( \ref{gfit}\right) $ \cite{Bornyakov:2003ee},
namely
\begin{equation}
m_{off}\approx 2m_{dg}\;,  \label{twice}
\end{equation}
were $m_{off}\;$has approximately the same value as that obtained in \cite
{Amemiya:1998jz}, $m_{off}\approx 1.2GeV$. Equation $\left( \ref{twice}%
\right) $ implies that the off-diagonal propagator is short-ranged as
compared to the diagonal one.\newline
\newline
Although the extrapolation of the lattice data in the region
$q\approx 0$ is a difficult task, which requires rather large
lattice volumes, our results on the transverse diagonal and
off-diagonal components of the gluon propagator can be considered
in qualitative agreement with the lattice results,
especially with the two parameter fits $\left( \ref{gfit}\right) $ and $%
\left( \ref{yfit}\right) $. Concerning now the ghost propagator, to our
knowledge, no lattice data are available so far.\newline
\newline
We remark here that the authors \cite{Bornyakov:2003ee} have also
reported a nonvanishing off-diagonal longitudinal component of the
gluon propagator which, in the low momentum region, seems to
behave in a way similar to the off-diagonal scalar function of
eq.$\left( \ref{yfit}\right) $. Nevertheless, the analytical
investigation of this issue would require a formulation which goes
beyond the original Gribov's quadratic approximation for the form
factor $\sigma(0,A)$, which has been employed in the present work,
see eqs.$\left( \ref{nps}\right)$,$\left( \ref{pf1}\right)$. This
approximation enables us to work out a first study of the
influence of the Gribov copies on the infrared behavior of the
gluon and ghost propagators. Moreover, the analysis is by no means
exhaustive and further work is certainly needed. In particular,
this approximation does not allow to take in due account quantum
corrections to the propagators in the presence of the Gribov
horizon. One should remark in fact that the longitudinal
off-diagonal propagator identically vanishes at the tree level, as
it is easily checked from the Feynman rules stemming from the
gauge fixing condition $D_{\mu}^{ab}A^{b}_{\mu}=0$. However, due
to the nonlinearity of the maximal Abelian gauge, one could argue
that a nonvanishing off-diagonal longitudinal propagator might
arise due to nonperturbative quantum effects. The transverse
diagonal and off-diagonal propagators, eq.$\left( \ref{gl}\right)
$ and eq.$\left( \ref{offg}\right)$, represent a kind of first
order propagators incorporating the effects of the Gribov horizon
as well
as of the dimension two condensate $%
\left\langle A_{\mu }^{a}A_{\mu }^{a}\right\rangle $. These
propagators have to be used in order to investigate higher order
quantum corrections as, for instance, the off-diagonal gluon
vacuum polarization which could give rise to a longitudinal
component of the off-diagonal propagator. Nevertheless, for  a
consistent evaluation of these quantum effects, we should have at
our disposal a local and renormalizable action which takes into
account the restriction to the Gribov region $\mathcal{C}_{0}$,
eq.$\left( \ref{gr}\right)$. The construction of such an action
has been achieved by Zwanziger \cite
{Zwanziger:1989mf,Zwanziger:1992qr} in the case of the Landau
gauge, where a suitable horizon function implementing the
restriction to the Gribov horizon has been identified. Remarkably,
the resulting action can be made local and enjoys the property of
being multiplicatively renormalizable. It can be effectively used
to evaluate quantum corrections by taking into account the
restriction to the first Gribov horizon, see for instance the
recent work \cite{Dudal:2005na}. Although being beyond the aim of
the present work, we mention that the study of the horizon
function for the maximal Abelian gauge is under investigation. Its
identification would allow us to properly address the issue of the
existence of a nonperturbative off-diagonal longitudinal gluon
propagator by analytical methods.

\section{Conclusion}

In this work the effects of the Gribov copies on the gluon and ghost
propagators in $SU(2)$ Euclidean Yang-Mills theory quantized in the maximal
Abelian gauge have been investigated. \newline
\newline
The domain of integration in the path integral has been restricted to the
Gribov region $\mathcal{C}_{0}$, defined as the set of field configurations
fulfilling the gauge conditions (\ref{offgauge}), (\ref{dgauge}), and for
which the Faddeev-Popov operator $\mathcal{M}^{ab}$, eq.(\ref{offop}), is
positive definite. Gribov's original statement \cite{Gribov:1977wm} about
closely related gauge copies located on opposite sides of a Gribov horizon
has been generalized to the maximal Abelian gauge, see Appendix \ref{A},
providing thus a support for the restriction of the domain of integration to
the region $\mathcal{C}_{0}$. The dimension two gluon condensate $%
\left\langle A_{\mu }^{a}A_{\mu }^{a}\right\rangle $ has also been taken
into account. \newline
\newline
The diagonal component of the gluon propagator displays a Gribov
type behavior in the infrared, eq.$\left( \ref{gl}\right) $. The
off-diagonal transverse component has been found to be of the
Yukawa type, with a dynamical gluon mass originating from
$\left\langle A_{\mu }^{a}A_{\mu }^{a}\right\rangle $, eq.$\left(
\ref{offg}\right) $. Moreover, the off-diagonal ghost propagator
exhibits infrared enhancement, eq.$\left( \ref{offgh}\right) $,
while the diagonal ghost propagator remains unaltered. Concerning
the behavior of the transverse diagonal and off-diagonal
components of the gluon propagator, our results can be considered
in qualitative agreement with those of lattice numerical
simulations \cite{Amemiya:1998jz,Bornyakov:2003ee}. \newline
\newline
Finally, we hope that this work will stimulate further investigation on the
behavior of the propagators in the maximal Abelian gauge from our colleagues
of the lattice community. A look at the off-diagonal ghost propagator would
be of a certain interest for a better understanding of the role of the
Gribov copies in this gauge.

\section*{Acknowledgments}

The Conselho Nacional de Desenvolvimento Cient\'{i}fico e Tecnol\'{o}gico
(CNPq-Brazil), the Faperj, Funda{\c{c}}{\~{a}}o de Amparo {\`{a}} Pesquisa
do Estado do Rio de Janeiro, the SR2-UERJ and the Coordena{\c{c}}{\~{a}}o de
Aperfei{\c{c}}oamento de Pessoal de N{\'{i}}vel Superior (CAPES) are
gratefully acknowledged for financial support.

\appendix


\section{A generalization of Gribov's statement to the maximal Abelian gauge %
\label{A}}

This Appendix is devoted to the generalization to the maximal Abelian gauge
of Gribov's statement \cite{Gribov:1977wm} about closely related copies
located on opposite sides of a Gribov horizon. Let us begin by reminding
that, as pointed out in \cite{Bruckmann:2000xd}, the Faddeev-Popov operator $%
\mathcal{M}^{ab}$
\begin{equation}
\mathcal{M}^{ab}(A)=-D_{\mu }^{ac}(A)D_{\mu }^{cb}(A)-g^{2}\varepsilon
^{ac}\varepsilon ^{bd}A_{\mu }^{c}A_{\mu }^{d}\;,  \label{fpaa}
\end{equation}
enjoys the property of being Hermitian, being the difference of two positive
semidefinite operators given, respectively, by $-D_{\mu }^{ac}D_{\mu }^{cb}$
and $g^{2}\varepsilon ^{ac}\varepsilon ^{bd}A_{\mu }^{c}A_{\mu }^{d}$. Its
eigenvalues are thus real. \newline
\newline
Following \cite{Gribov:1977wm}, we can divide the space of fields fulfilling
the gauge conditions (\ref{offgauge}) and (\ref{dgauge}) into regions with a
definite number of bound states, \textit{i.e.} negative energy solutions of
the operator $\mathcal{M}^{ab}$, see Fig.1.
\begin{figure}[th]
\begin{center}
\scalebox{0.7}{\includegraphics{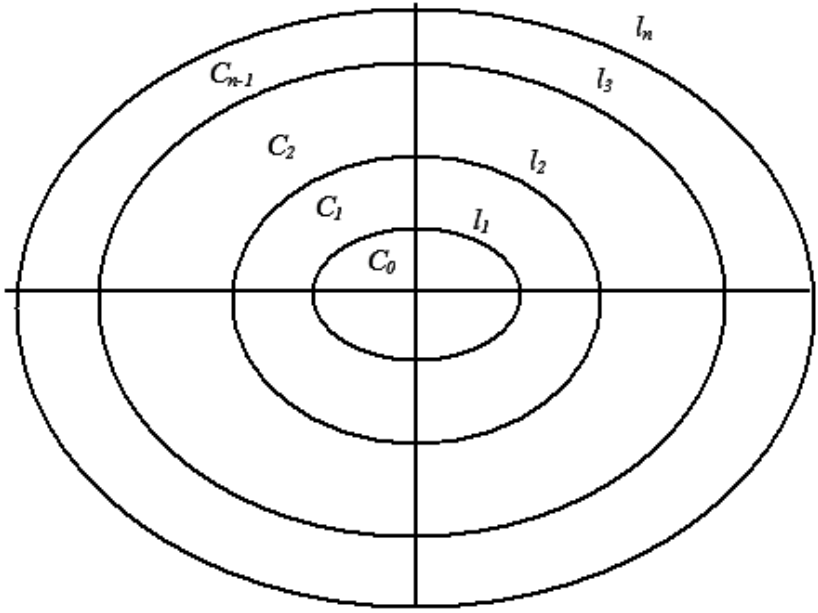}}
\end{center}
\caption{The Gribov horizons }
\end{figure}

\noindent Let us look thus at the eigenvalues equation for the Faddeev-Popov
operator $\mathcal{M}^{ab}$, \textit{i.e.}
\begin{equation}
\mathcal{M}^{ab}\psi ^{b}=\epsilon (A)\psi ^{a}\;.  \label{eig}
\end{equation}
For small values of the gauge fields $(A_{\mu },A_{\mu }^{a})$, eq.$\left(
\ref{eig}\right) $ is solvable for positive $\epsilon (A)$ only. More
precisely, denoting by $\epsilon _{1}(A),\epsilon _{2}(A),\epsilon
_{3}(A),....,$ the eigenvalues corresponding to a given field configuration $%
(A_{\mu },A_{\mu }^{a})$, one has that, for small $(A_{\mu },A_{\mu }^{a})$,
all $\epsilon _{i}(A)$ are positive, $\epsilon _{i}(A)>0$, corresponding to
field configurations for which $-D_{\mu }^{ac}D_{\mu }^{cb}>g^{2}\varepsilon
^{ac}\varepsilon ^{bd}A_{\mu }^{c}A_{\mu }^{d}$. However, for a sufficiently
large value of the fields $(A_{\mu },A_{\mu }^{a})$, one of the eigenvalues,
say $\epsilon _{1}(A)$, turns out to vanish, becoming negative as the fields
increase further\footnote{%
See also the argument presented in Sect.3 of \cite{Bruckmann:2000xd}.}. This
means that the fields $(A_{\mu },A_{\mu }^{a})$ are large enough to ensure
the existence of negative energy solutions, \textit{i.e.} bound states. For
a greater magnitude of $(A_{\mu },A_{\mu }^{a})$, a second eigenvalue, say $%
\epsilon _{2}(A)$, will vanish, becoming negative as the fields increase
again. Following Gribov \cite{Gribov:1977wm}, we may thus divide the
functional space of the fields into regions $\mathcal{C}_{0},\mathcal{C}_{1},%
\mathcal{C}_{2},...,\mathcal{C}_{n}$ over which the operator $\mathcal{M}%
^{ab}$ has $0,1,2,....,n$ negative eigenvalues. These regions are separated
by lines $l_{1},l_{2},l_{3},...,l_{n}$ on which the operator $\mathcal{M}%
^{ab}$ has zero energy solutions. The meaning of Fig.1 is as follows. In the
region $\mathcal{C}_{0}$ all eigenvalues of the operator $\mathcal{M}^{ab}$
are positive, \textit{i.e.} $\mathcal{M}^{ab}>0$. At the boundary $l_{1}$ of
the region $\mathcal{C}_{0}$ the first vanishing eigenvalue appears, namely
on $l_{1}$ the operator $\mathcal{M}^{ab}$ possesses a normalizable zero
mode. In the region $\mathcal{C}_{1}$ the operator $\mathcal{M}^{ab}$ has
one bound state, \textit{\ i.e.} one negative energy solution. At the
boundary $l_{2}$, a zero eigenvalue reappears. In the region $\mathcal{C}_{2}
$ the operator $\mathcal{M}^{ab}$ has two bound states, \textit{i.e.} two
negative energy solutions. On $l_{3}$ a zero eigenvalue shows up again, and
so on. The boundaries $l_{1},\;l_{2},\;l_{3},\;....,\;l_{n}$,$\;$ on which
the operator $\mathcal{M}^{ab}$ has zero eigenvalues are called Gribov
horizons. In particular, the boundary $l_{1}$ where the first vanishing
eigenvalue appears is called the first horizon. See \cite{Bruckmann:2000xd}
for an explicit example of a horizon configuration. \newline
\newline
It is useful to emphasize that in the region $\mathcal{C}_{0}$, the operator
$\mathcal{M}^{ab}$ has only positive eigenvalues. Therefore, this region can
be defined as the set of all gauge fields $(A_{\mu },A_{\mu }^{a})$
fulfilling the gauge conditions eqs.(\ref{offgauge}), (\ref{dgauge}), for
which the Faddeev-Popov operator $\mathcal{M}^{ab}$ is positive definite,
see eq.(\ref{gr}). Note also that field configurations belonging to $%
\mathcal{C}_{0}$ correspond to relative minima of the auxiliary functional $%
\mathcal{R}[A]$. This follows by observing that the Faddeev-Popov operator $%
\mathcal{M}^{ab}$ can be obtained by taking the second variation of $%
\mathcal{R}[A]$ \cite{Bruckmann:2000xd}. \newline
\newline
Let us proceed with the generalization to the maximal Abelian gauge of
Gribov's result stating that for any field close to a horizon there is an
equivalent field, \textit{i.e.} a gauge copy, located on the other side of
the horizon, close to the same horizon, see Fig.2
\begin{figure}[th]
\begin{center}
\scalebox{0.7}{\includegraphics{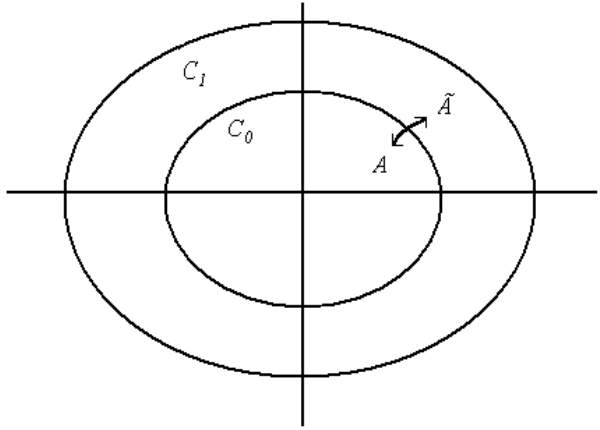}}
\end{center}
\caption{The equivalent fields }
\end{figure}

\vspace{1cm} \noindent Let us start by considering a field configuration $%
\left( C_{\mu },\;C_{\mu }^{a}\right) $ located on the first Gribov horizon $%
l_{1}$, namely
\begin{eqnarray}
\mathcal{M}^{ab}\left( C\right) \varphi _{0}^{b} &=&-\left( D_{\mu
}^{ac}(C)D_{\mu }^{cb}(C)+g^{2}\varepsilon ^{ac}\varepsilon ^{bd}C_{\mu
}^{c}C_{\mu }^{d}\right) \varphi _{0}^{b}=0\;,  \nonumber \\
D_{\mu }^{ab}\left( C_{\mu }\right) C_{\mu }^{b} &=&0\;,  \nonumber \\
\partial _{\mu }C_{\mu } &=&0\;,  \label{hor0}
\end{eqnarray}
where $\varphi _{0}^{a}$ denotes a normalizable zero mode. In the following
it turns out to be useful to introduce the diagonal component $\varphi _{0}$
which, according to eq.(\ref{des}), is defined as
\begin{equation}
\varphi _{0}=-g\epsilon ^{ab}\frac{\partial _{\mu }}{\partial ^{2}}\left(
C_{\mu }^{a}\varphi _{0}^{b}\right) \;.
\end{equation}
Let thus $\left( A_{\mu },A_{\mu }^{a}\right) $ be a field configuration
located in the Gribov region $\mathcal{C}_{0}$, close to the horizon $l_{1}$%
, Fig.2. Following \cite{Gribov:1977wm} we write
\begin{eqnarray}
A_{\mu }^{a} &=&C_{\mu }^{a}+a_{\mu }^{a}\;,  \nonumber \\
A_{\mu } &=&C_{\mu }+a_{\mu }\;,  \label{near}
\end{eqnarray}
where $\left( a_{\mu },\;a_{\mu }^{a}\right) $ have to be considered as
small perturbations. The fields $(A_{\mu },A_{\mu }^{a})$ obey the gauge
conditions (\ref{offgauge}) and  (\ref{dgauge}) which, neglecting higher
order terms in the small components $\left( a_{\mu },\;a_{\mu }^{a}\right) $%
, read
\begin{eqnarray*}
D_{\mu }^{ab}(C)a_{\mu }^{b}-g\varepsilon ^{ab}C_{\mu }^{b}a_{\mu } &=&0\;,
\\
\partial _{\mu }a_{\mu } &=&0\;.
\end{eqnarray*}
The evaluation of the energy eigenvalue $\epsilon (A)$ of the Faddeev-Popov
operator $\mathcal{M}^{ab}(A)$ corresponding to the field configuration $%
\left( A_{\mu },A_{\mu }^{a}\right) $ can be easily handled by means of
perturbation theory, yielding
\begin{equation}
\epsilon (A)=\frac{\int d^{4}x\;\varphi _{0}^{a}\left( 2g\varepsilon
^{ac}a_{\mu }D_{\mu }^{cb}(C)\varphi _{0}^{b}+g^{2}\varepsilon
^{ac}\varepsilon ^{db}(C_{\mu }^{c}a_{\mu }^{d}+C_{\mu }^{d}a_{\mu
}^{c})\varphi _{0}^{b}\right) }{\int d^{4}x\;\varphi _{0}^{a}\varphi _{0}^{a}%
}\;,  \label{ea}
\end{equation}
Proceeding as in \cite{Gribov:1977wm}, we introduce the fields
\begin{eqnarray}
\tilde{A}_{\mu }^{a} &=&C_{\mu }^{a}+\tilde{a}_{\mu }^{a}\;,  \nonumber \\
\tilde{A}_{\mu } &=&C_{\mu }+\tilde{a}_{\mu }\;,  \label{tfield}
\end{eqnarray}
where
\begin{eqnarray}
\tilde{a}_{\mu }^{a} &=&a_{\mu }^{a}-D_{\mu }^{ab}(C)\varphi
_{0}^{b}-g\varepsilon ^{ab}C_{\mu }^{b}\varphi _{0}\;,  \nonumber \\
\tilde{a}_{\mu } &=&a_{\mu }-\partial _{\mu }\varphi _{0}-g\varepsilon
^{ab}C_{\mu }^{a}\varphi _{0}^{b}\;,  \label{taf}
\end{eqnarray}
have to be considered as small as compared to $(C_{\mu },C_{\mu }^{a})$. It
is not difficult to verify that, to first order in the small components $%
\left( \tilde{a}_{\mu }^{a},\tilde{a}_{\mu }\right) $, the fields $(\tilde{A}%
_{\mu },\tilde{A}_{\mu }^{a})$ obey the same gauge conditions of $(A_{\mu
},A_{\mu }^{a})$, namely
\begin{eqnarray}
D_{\mu }^{ab}(\tilde{A})\tilde{A}_{\mu }^{b} &=&0\;,  \nonumber \\
\partial _{\mu }\tilde{A}_{\mu } &=&0\;.  \label{req}
\end{eqnarray}
The fields $(\tilde{A}_{\mu },\tilde{A}_{\mu }^{a})$ might thus be
identified with a Gribov copy of $(A_{\mu },A_{\mu }^{a})$, provided one is
able to find a gauge transformation $S$ such that
\begin{eqnarray}
\widetilde{\mathcal{A}}_{\mu } &=&S^{\dagger }\mathcal{{A}_{\mu }}%
S+S^{\dagger }\partial _{\mu }S\;,  \nonumber \\
\widetilde{\mathcal{A}}_{\mu } &=&\tilde{A_{\mu }^{a}}T^{a}+\tilde{A_{\mu }}%
T^{3}\;,  \nonumber \\
\mathcal{A}_{\mu } &=&A_{\mu }^{a}T^{a}+A_{\mu }T^{3}\;.  \label{tg}
\end{eqnarray}
We shall look at $S$ close to unit, in the form
\begin{eqnarray}
S &=&1-\overline{\alpha }+\frac{\overline{\alpha }^{2}}{2}+O(\overline{%
\alpha }^{3})\;,  \nonumber \\
\overline{\alpha } &=&\alpha ^{a}T^{a}+\alpha T^{3}\;,  \label{gex}
\end{eqnarray}
from which we obtain
\begin{eqnarray}
\tilde{A}_{\mu }^{a} &=&A_{\mu }^{a}-\left( D_{\mu }^{ab}\alpha
^{b}+g\varepsilon ^{ab}A_{\mu }^{b}\alpha \right) -\frac{g}{2}\varepsilon
^{ab}\alpha ^{b}\left( \partial _{\mu }\alpha +g\varepsilon ^{cd}A_{\mu
}^{c}\alpha ^{d}\right) +\frac{g}{2}\varepsilon ^{ab}\alpha {D}_{\mu
}^{bc}\alpha ^{c}-\frac{g^{2}}{2}A_{\mu }^{a}\alpha ^{2}\;,  \nonumber \\
\tilde{A}_{\mu } &=&A_{\mu }-\left( \partial _{\mu }\alpha +g\varepsilon
^{ab}A_{\mu }^{a}\alpha ^{b}\right) -\frac{g}{2}\varepsilon ^{ab}\alpha
^{a}D_{\mu }^{bc}\alpha ^{c}+\frac{g^{2}}{2}A_{\mu }^{a}\alpha ^{a}\alpha \;.
\nonumber \\
&&  \label{2ordemcopias}
\end{eqnarray}
Furthermore, from eq.(\ref{req}), it follows
\begin{eqnarray}
\mathcal{M}^{ab}(A)\alpha ^{b}+D_{\mu }^{ab}(A)\left[ -\frac{g}{2}%
\varepsilon ^{bc}\alpha ^{c}\left( \partial _{\mu }\alpha +g\varepsilon
^{de}A_{\mu }^{d}\alpha ^{e}\right) +\frac{g}{2}\varepsilon ^{bc}\alpha {D}%
_{\mu }^{cd}\alpha ^{d}-\frac{g^{2}}{2}A_{\mu }^{b}\alpha ^{2}\right]  &{\ }&
\nonumber \\
-g\varepsilon ^{ab}\left( \partial _{\mu }\alpha +g\varepsilon ^{cd}A_{\mu
}^{c}\alpha ^{d}\right) \left( D_{\mu }^{be}\alpha ^{e}+g\varepsilon
^{be}A_{\mu }^{e}\alpha \right) +g\varepsilon ^{ab}A_{\mu }^{b}\left( \frac{g%
}{2}\varepsilon ^{cd}\alpha ^{c}D_{\mu }^{de}\alpha ^{e}-\frac{g^{2}}{2}%
A_{\mu }^{c}\alpha ^{c}\alpha \right)  &=&0\;,  \nonumber \\[4mm]
\partial _{\mu }\left[ -\left( \partial _{\mu }\alpha +g\varepsilon
^{ab}A_{\mu }^{a}\alpha ^{b}\right) -\frac{g}{2}\varepsilon ^{ab}\alpha
^{a}D_{\mu }^{bc}\alpha ^{c}+\frac{g^{2}}{2}A_{\mu }^{a}\alpha ^{a}\alpha
\right]  &=&0\;.  \nonumber \\
&&  \label{2ordemcopias1}
\end{eqnarray}
In order to express $\left( \alpha ,\alpha ^{a}\right) $ in terms of $\left(
\varphi _{0},\varphi _{0}^{a}\right) $, we follow \cite{Gribov:1977wm}, and
set
\begin{eqnarray}
\alpha ^{a} &=&\varphi _{0}^{a}+\tilde{\varphi}^{a}\;,  \nonumber \\
\alpha  &=&\varphi _{0}+\tilde{\varphi}\;,  \label{gaugepar}
\end{eqnarray}
with $\left( \tilde{\varphi},\tilde{\varphi}^{a}\right) \;$small with
respect to $\left( \varphi _{0},\varphi _{0}^{a}\right) $. Condition (\ref
{2ordemcopias1}) gives thus
\begin{eqnarray}
\mathcal{M}^{ab}(C)\tilde{\varphi}^{b} &=&-g\varepsilon ^{cb}D_{\mu
}^{ac}(C)(a_{\mu }\varphi _{0}^{b})-g\varepsilon ^{ac}a_{\mu }D_{\mu
}^{cb}(C)\varphi _{0}^{b}-g^{2}\varepsilon ^{ac}\varepsilon ^{db}(C_{\mu
}^{c}a_{\mu }^{d}+C_{\mu }^{d}a_{\mu }^{c})\varphi _{0}^{b}  \nonumber \\
&+&D_{\mu }^{ab}(C)\left[ \frac{g}{2}\varepsilon ^{bc}\varphi
_{0}^{c}(\partial _{\mu }\varphi _{0}+g\varepsilon ^{de}C_{\mu }^{d}\varphi
_{0}^{e})-\frac{g}{2}\varepsilon ^{bc}\varphi _{0}D_{\mu }^{cd}(C)\varphi
_{0}^{d}+\frac{g^{2}}{2}C_{\mu }^{b}\varphi _{0}^{2}\right]   \nonumber \\
&+&g\varepsilon ^{ab}(\partial _{\mu }\varphi _{0}+g\varepsilon ^{cd}C_{\mu
}^{c}\varphi _{0}^{d})\left[ D_{\mu }^{be}(C)\varphi _{0}^{e}+g\varepsilon
^{be}C_{\mu }^{e}\varphi _{0}\right]   \nonumber \\
&-&g\varepsilon ^{ab}C_{\mu }^{b}\left[ \frac{g}{2}\varepsilon ^{cd}\varphi
_{0}^{c}D_{\mu }^{de}(C)\varphi _{0}^{e}-\frac{g^{2}}{2}C_{\mu }^{c}\varphi
_{0}^{c}\varphi _{0}\right] \;.  \nonumber \\
&&  \label{2ordemcopias2}
\end{eqnarray}
Note that eq.(\ref{2ordemcopias2}) can be cast in the form
\begin{equation}
\partial ^{2}\tilde{\varphi}^{a}=\mathcal{P}^{a}(C,a,\varphi _{0})+\mathcal{Q%
}^{ab}(C,\varphi _{0})\tilde{\varphi}^{b}\;,  \label{cf1}
\end{equation}
where $\mathcal{P}^{a}$ and $\mathcal{Q}^{ab}$ are independent from $\tilde{%
\varphi}$, \textit{i.e.}
\begin{eqnarray}
\mathcal{-P}^{a}(C,a,\varphi _{0}) &=&-g\varepsilon ^{cb}D_{\mu
}^{ac}(C)(a_{\mu }\varphi _{0}^{b})-g\varepsilon ^{ac}a_{\mu }D_{\mu
}^{cb}(C)\varphi _{0}^{b}-g^{2}\varepsilon ^{ac}\varepsilon ^{db}(C_{\mu
}^{c}a_{\mu }^{d}+C_{\mu }^{d}a_{\mu }^{c})\varphi _{0}^{b}  \nonumber \\
&+&D_{\mu }^{ab}(C)\left[ \frac{g}{2}\varepsilon ^{bc}\varphi
_{0}^{c}(\partial _{\mu }\varphi _{0}+g\varepsilon ^{de}C_{\mu }^{d}\varphi
_{0}^{e})-\frac{g}{2}\varepsilon ^{bc}\varphi _{0}D_{\mu }^{cd}(C)\varphi
_{0}^{d}+\frac{g^{2}}{2}C_{\mu }^{b}\varphi _{0}^{2}\right]   \nonumber \\
&+&g\varepsilon ^{ab}(\partial _{\mu }\varphi _{0}+g\varepsilon ^{cd}C_{\mu
}^{c}\varphi _{0}^{d})\left[ D_{\mu }^{be}(C)\varphi _{0}^{e}+g\varepsilon
^{be}C_{\mu }^{e}\varphi _{0}\right]   \nonumber \\
&-&g\varepsilon ^{ab}C_{\mu }^{b}\left[ \frac{g}{2}\varepsilon ^{cd}\varphi
_{0}^{c}D_{\mu }^{de}(C)\varphi _{0}^{e}-\frac{g^{2}}{2}C_{\mu }^{c}\varphi
_{0}^{c}\varphi _{0}\right] \;,  \nonumber \\
&&  \label{cf2}
\end{eqnarray}
and
\begin{equation}
\mathcal{Q}^{ab}(C,\varphi _{0})=2g\varepsilon ^{ab}C_{\mu }\partial _{\mu
}-g^{2}\varepsilon ^{ac}\varepsilon ^{cb}C_{\mu }C_{\mu }-g^{2}\varepsilon
^{ac}\varepsilon ^{bd}C_{\mu }^{c}C_{\mu }^{d}\;.  \label{cf3}
\end{equation}
Equation (\ref{cf1}) can be solved recursively for
$\tilde{\varphi}^{a}$, namely
\begin{equation}
\tilde{\varphi}^{a}=\frac{1}{\partial ^{2}}\mathcal{P}^{a}+\frac{1}{\partial
^{2}}\mathcal{Q}^{ab}\frac{1}{\partial ^{2}}\mathcal{P}^{b}+......\;
\label{cf4}
\end{equation}
This allows us to obtain a recursive expression for the parameters
$\left( \alpha ,\alpha ^{a}\right) $ of eq.(\ref{gaugepar}), an
thus for the gauge transforamtion $S$ we are looking for,
eq.(\ref{gex}). \\\\Moreover, recalling that
\begin{equation}
\mathcal{M}^{ab}\left( C\right) \varphi _{0}^{b}=0\;,  \label{rec}
\end{equation}
one finds
\begin{equation}
\int d^{4}x\;\varphi _{0}^{a}\mathcal{M}^{ab}(C)\tilde{\varphi}^{b}\;=0\;,
\label{cd1}
\end{equation}
so that

\begin{eqnarray}
0 &=&\int d^{4}x\;\varphi _{0}^{a}\;\left[ -2g\varepsilon ^{ac}a_{\mu
}D_{\mu }^{cb}(C)\varphi _{0}^{b}-g^{2}\varepsilon ^{ac}\varepsilon
^{db}(C_{\mu }^{c}a_{\mu }^{d}+C_{\mu }^{d}a_{\mu }^{c})\varphi
_{0}^{b}\right.  \nonumber \\
&&\;\;\left. \;+g\varepsilon ^{ab}(\partial _{\mu }\varphi _{0})D_{\mu
}^{bc}(C)\varphi _{0}^{c}-2g^{2}\varepsilon ^{ab}\varepsilon ^{cd}C_{\mu
}^{b}\varphi _{0}^{c}D_{\mu }^{de}(C)\varphi _{0}^{e}+g^{3}\varepsilon
^{ab}C_{\mu }^{b}C_{\mu }^{c}\varphi _{0}^{c}\varphi _{0}\right] \;.
\label{cd2}
\end{eqnarray}
Let us now check on which side of the horizon $l_{1}$ the equivalent fields $%
(\tilde{A}_{\mu },\tilde{A}_{\mu }^{a})$ are located. As done before, we
look at the energy eigenvalue $\epsilon (\tilde{A})$, given by

\begin{equation}
\epsilon (\tilde{A})=\frac{\int d^{4}x\;\varphi _{0}^{a}\left( 2g\varepsilon
^{ac}\tilde{a}_{\mu }D_{\mu }^{cb}(C)\varphi _{0}^{b}+g^{2}\varepsilon
^{ac}\varepsilon ^{db}(C_{\mu }^{c}\tilde{a}_{\mu }^{d}+C_{\mu }^{d}\tilde{a}%
_{\mu }^{c})\varphi _{0}^{b}\right) }{\int d^{4}x\;\varphi _{0}^{a}\varphi
_{0}^{a}}\;.  \label{eiat}
\end{equation}
Finally, form eqs.(\ref{taf}), (\ref{cd2}) it follows that
\begin{equation}
\epsilon (\tilde{A})=-\epsilon (A)\;.  \label{fe}
\end{equation}
Thus, if the configuration $(A_{\mu },A_{\mu }^{a})$, close to $l_{1}$, is
located in the region $\mathcal{C}_{0}$, $\epsilon (A)>0$, there is an
equivalent field configuration $(\tilde{A}_{\mu },\tilde{A}_{\mu }^{a})$,
eqs.(\ref{tfield}), (\ref{taf}), close to $l_{1}$, which is located in $%
\mathcal{C}_{1}$, $\epsilon (\tilde{A})=-\epsilon (A)<0$. This derivation,
which can be repeated to fields close to any horizon $l_{n}$, generalizes
Gribov's statement to the case of the maximal Abelian gauge .


\section{Faddeev-Popov quantization of Yang-Mills theory in the maximal
Abelian gauge \label{B}}

We provide here a detailed summary of the Faddeev-Popov quantization of
Yang-Mills theory in the maximal Abelian gauge. Following \cite{Dudal:2004rx}%
, let us start by giving the $BRST$ transformations of all fields, namely
\label{B}
\begin{eqnarray}
sA_{\mu }^{a} &=&-\left( D_{\mu }^{ab}c^{b}+g\varepsilon ^{\,ab}A_{\mu
}^{b}c\right) \;,  \nonumber \\
sA_{\mu } &=&-\left( \partial _{\mu }c+g\varepsilon ^{ab}A_{\mu
}^{a}c^{b}\right) \;,  \nonumber \\
sc^{a} &=&g\varepsilon \,^{ab}c^{b}c\;,  \nonumber \\
\,sc &=&\frac{g}{2}\,\varepsilon \,^{ab}c^{a}c^{b},  \nonumber \\
s\overline{c}^{a} &=&b^{a}\;,  \nonumber \\
s\overline{c} &=&b\;,  \nonumber \\
sb^{a} &=&0\;,  \nonumber \\
\,sb &=&0\;,  \label{brs}
\end{eqnarray}
with
\begin{equation}
s^{2}=0\;,  \label{nbrs}
\end{equation}
where $\left( \overline{c}^{a},c^{a}\right) $ and $\left( \overline{c}%
,c\right) $ are the off-diagonal and diagonal Faddeev-Popov ghosts, while $%
\left( b^{a},b\right) \;$denote the Lagrange multipliers. For the
gauge-fixed Yang-Mills theory one has
\begin{equation}
S_{\mathrm{YM}}+S_{\mathrm{MAG}}+S_{\mathrm{diag}}\;,  \label{smag}
\end{equation}
\begin{equation}
s\left( S_{\mathrm{YM}}+S_{\mathrm{MAG}}+S_{\mathrm{diag}}\right) =0\;,
\label{sinv}
\end{equation}
where
\begin{equation}
S_{\mathrm{YM}}=\frac{1}{4}\int d^{4}x\,\left( F_{\mu \nu }^{a}F_{\mu \nu
}^{a}+F_{\mu \nu }F_{\mu \nu }\right) \;,  \label{ymmag}
\end{equation}
and $S_{\mathrm{MAG}},$\ $S_{\mathrm{diag}}$ being the gauge fixing terms
corresponding to the off-diagonal and diagonal sectors, respectively. They
are given by
\begin{eqnarray}
S_{\mathrm{MAG}} &=&s\,\int d^{4}x\,\left( \overline{c}^{a}\left( D_{\mu
}^{ab}A_{\mu }^{b}+\frac{\alpha }{2}b^{a}\right) -\frac{\alpha }{2}%
g\varepsilon \,^{ab}\overline{c}^{a}\overline{c}^{b}c\right) \;  \nonumber \\
&=&\int d^{4}x\left( b^{a}\left( D_{\mu }^{ab}A_{\mu }^{b}+\frac{\alpha }{2}%
b^{a}\right) +\overline{c}^{a}D_{\mu }^{ab}D_{\mu }^{bc}c^{c}+g\overline{c}%
^{a}\varepsilon ^{ab}\left( D_{\mu }^{bc}A_{\mu }^{c}\right) c\right.
\nonumber \\
&&\left. -\alpha g\varepsilon ^{ab}b^{a}\overline{c}^{b}c-g^{2}\varepsilon
^{ab}\varepsilon ^{cd}\overline{c}^{a}c^{d}A_{\mu }^{b}A_{\mu }^{c}-\frac{%
\alpha }{4}g^{2}\varepsilon ^{ab}\varepsilon ^{cd}\overline{c}^{a}\overline{c%
}^{b}c^{c}c^{d}\right) \;,  \label{smm}
\end{eqnarray}
and
\begin{equation}
S_{\mathrm{diag}}=s\,\int d^{4}x\,\;\overline{c}\partial _{\mu }A_{\mu
}\;=\int d^{4}x\,\;\left( b\partial _{\mu }A_{\mu }+\overline{c}\partial
_{\mu }\left( \partial _{\mu }c+g\varepsilon ^{ab}A_{\mu }^{a}c^{b}\right)
\right) \;.  \label{sgfd}
\end{equation}
Note that, for renormalizability purposes, a gauge parameter $\alpha $ has
to be introduced in the off-diagonal part of the gauge fixing, eq.$\left(
\ref{smm}\right) $. The maximal Abelian gauge condition, $D_{\mu
}^{ab}A_{\mu }^{b}=0$, is recovered in the limit $\alpha \rightarrow $ $0$,
which has to be taken after the removal of the ultraviolet divergences \cite
{Dudal:2004rx}. In fact, some of the terms proportional to $\alpha $ would
reappear due to radiative corrections, even if $\alpha =0$. See, for
example, \cite{Kondo:2001tm}. Moreover, the action $\left( \ref{smag}\right)
$ is multiplicatively renormalizable to all orders of perturbation theory
\cite{Fazio:2001rm,Dudal:2004rx}. \newline
\newline
Therefore, for the partition function expressing the Faddeev-Popov
quantization of Yang-Mills theory in the maximal Abelian gauge we have
\begin{equation}
\mathcal{Z}=\int DA_{\mu }DA_{\mu }^{a}Db^{a}DbDc^{a}DcD\overline{c}^{a}D%
\overline{c}\;e^{-\left( S_{\mathrm{YM}}+S_{\mathrm{MAG}}+S_{\mathrm{diag}%
}\right) \label{zzz} \;}.
\end{equation}
Taking the limit $\alpha \rightarrow $ $0$ and integrating over the Lagrange
multipliers $\left( b^{a},b\right) $, one gets
\begin{eqnarray}
\mathcal{Z} &=&\int DA_{\mu }DA_{\mu }^{a}Dc^{a}DcD\overline{c}^{a}D%
\overline{c}\;\delta \left( D_{\mu }^{ab}A_{\mu }^{b}\right) \delta \left(
\partial _{\mu }A_{\mu }\right) e^{-S_{\mathrm{YM}}\;}  \nonumber \\
&&\;\;\;\;\;\;\times \exp \left( -\int d^{4}x\left( \,\left( \overline{c}%
^{a}D_{\mu }^{ab}D_{\mu }^{bc}c^{c}-g^{2}\varepsilon ^{ab}\varepsilon ^{cd}%
\overline{c}^{a}c^{d}A_{\mu }^{b}A_{\mu }^{c}\right) +\overline{c}\partial
_{\mu }\left( \partial _{\mu }c+g\varepsilon ^{ab}A_{\mu }^{a}c^{b}\right)
\right) \right) \;.  \nonumber \\
&&  \label{z1}
\end{eqnarray}
To deal with the diagonal ghosts $(\overline{c},c)$ we perform now the
change of variables
\begin{eqnarray}
c &\rightarrow &\widetilde{c}=c+g\frac{\partial _{\mu }}{\partial ^{2}}%
\left( \varepsilon ^{ab}A_{\mu }^{a}c^{b}\right) \;,  \nonumber \\
c^{a} &\rightarrow &c^{a}\;,  \nonumber \\
A_{\mu }^{a} &\rightarrow &A_{\mu }^{a}\;,  \label{chv}
\end{eqnarray}
all other fields remaining unchanged. It is easy to check that
\begin{equation}
\partial _{\mu }\left( \partial _{\mu }c+g\varepsilon ^{ab}A_{\mu
}^{a}c^{b}\right) \rightarrow \partial ^{2}\widetilde{c}\;,  \label{chv1}
\end{equation}
and that the Jacobian $J$ corresponding to $\left( \text{\ref{chv}}\right) $
is field independent. In fact

\begin{equation}
J=\det \left( \;
\begin{tabular}{lll}
$1$ & $g\varepsilon ^{ac}\frac{\partial _{\mu }}{\partial ^{2}}c^{c}$ & $%
g\varepsilon ^{cd}\frac{\partial _{\mu }}{\partial ^{2}}A_{\mu }^{c}$ \\
$0$ & $\delta ^{ab}\delta _{\mu \nu }$ & $0$ \\
$0$ & $0$ & $\delta ^{bd}$%
\end{tabular}
\right) =\mathrm{const}.\;\;  \label{jac}
\end{equation}
One sees thus that the transformation $\left( \text{\ref{chv}}\right) $
allows us to decouple the diagonal ghosts from the theory, namely
\begin{eqnarray}
\mathcal{Z} &=&\int DA_{\mu }DA_{\mu }^{a}Dc^{a}D\widetilde{c}D\overline{c}%
^{a}D\overline{c}\;\delta \left( D_{\mu }^{ab}A_{\mu }^{b}\right) \delta
\left( \partial _{\mu }A_{\mu }\right) e^{-S_{\mathrm{YM}}\;}  \nonumber \\
&&\;\;\;\;\;\;\times \exp \left( -\int d^{4}x\left( \,\left( \overline{c}%
^{a}D_{\mu }^{ab}D_{\mu }^{bc}c^{c}-g^{2}\varepsilon ^{ab}\varepsilon ^{cd}%
\overline{c}^{a}c^{d}A_{\mu }^{b}A_{\mu }^{c}\right) +\overline{c}\partial
^{2}\widetilde{c}\right) \right) \;,  \label{z2}
\end{eqnarray}
so that they can be integrated out, yielding
\begin{eqnarray}
\mathcal{Z} &=&\mathcal{N}\int DA_{\mu }DA_{\mu }^{a}Dc^{a}D\overline{c}%
^{a}\;\delta \left( D_{\mu }^{ab}A_{\mu }^{b}\right) \delta \left( \partial
_{\mu }A_{\mu }\right) e^{-S_{\mathrm{YM}}\;}  \nonumber \\
&&\;\;\;\;\;\;\times \exp \left( -\int d^{4}x\left( \,\left( \overline{c}%
^{a}D_{\mu }^{ab}D_{\mu }^{bc}c^{c}-g^{2}\varepsilon ^{ab}\varepsilon ^{cd}%
\overline{c}^{a}c^{d}A_{\mu }^{b}A_{\mu }^{c}\right) \right) \right) \;,
\label{ptqf}
\end{eqnarray}
where $\mathcal{N}$ is an irrelevant constant factor. It is worth remarking
that the change of variables $\left( \text{\ref{chv}}\right) $ seems to be a
peculiar feature of the maximal Abelian gauge. One could try in fact to
perform such a kind of transformation to decouple the Faddeev-Popov ghosts
in other cases as, for instance, the Landau gauge. However, it is
straightforward to check now that the Jacobian of the decoupling
transformation is no more field independent. It gives back precisely the
Faddeev-Popov determinant for the Landau gauge, so that the Faddeev-Popov
ghosts show up again. \newline
\newline
Finally, integrating over the off-diagonal ghosts $\left( \overline{c}%
^{a},c^{a}\right) $, it follows
\begin{equation}
\mathcal{Z}=\mathcal{N}\int DA_{\mu }DA_{\mu }^{a}\;\delta \left( D_{\mu
}^{ab}A_{\mu }^{b}\right) \delta \left( \partial _{\mu }A_{\mu }\right)
\left( \det \left( \mathcal{M}^{ab}\right) \right) e^{-S_{\mathrm{YM}}\;}\;,
\label{pftf}
\end{equation}
where $\mathcal{M}^{ab}$ is the off-diagonal Faddeev-Popov operator, as
given in eq.$\left( \text{\ref{offop}}\right) $. Expression $\left( \text{%
\ref{pftf}}\right) $ will be taken as the starting point for the
implementation of the restriction to the Gribov region $\mathcal{C}_{0}$ for
the maximal Abelian gauge.

\end{document}